\documentclass[12pt, twocolumn]{aastex631}

\usepackage{graphicx}

\shorttitle{Young Substellar Objects in IC\,348 and Barnard\,5}
\shortauthors{Lalchand et al.}
\graphicspath{{./}{figures/}}

\begin{document}

\title{A Novel Survey for Young Substellar Objects with the W-band Filter.V. IC\,348 and Barnard~5 in the Perseus Cloud}


\author[0000-0003-1618-2921]{Bhavana Lalchand}
\affiliation{Institute of Astronomy, National Central University, 300 Zhongda Road, Zhongli, Taoyuan 32001, Taiwan}

\author[0000-0003-0262-272X]{Wen-Ping Chen}
\affiliation{Institute of Astronomy, National Central University, 300 Zhongda Road, Zhongli, Taoyuan 32001, Taiwan}
\affiliation{Department of Physics,  
National Central University, 300 Zhongda Road, Zhongli, Taoyuan 32001, Taiwan}

\author[0000-0003-4614-7035]{Beth A. Biller}
\affiliation{SUPA, Institute for Astronomy, University of Edinburgh, Blackford Hill, Edinburgh EH9 3HJ, UK}
\affiliation{Centre for Exoplanet Science, University of Edinburgh, Edinburgh, UK}

\author{Lo\"{i}c Albert}
\affiliation{Institut de Recherche sur les Exoplan\`{e}tes (iREx), Universit\'{e} de Montr\'{e}al, D\'{e}partement de Physique, C.P. 6128 Succ. Centre-ville, Montr\'{e}al, QC H3C 3J7, Canada}

\author{Katelyn Allers}
\affiliation{Department of Physics and Astronomy, Bucknell University, Lewisburg, PA 17837, USA}

\author[0000-0001-6951-469X]{Sophie Dubber}
\affiliation{SUPA, Institute for Astronomy, University of Edinburgh, Blackford Hill, Edinburgh EH9 3HJ, UK}
\affiliation{Centre for Exoplanet Science, University of Edinburgh, Edinburgh, UK}

\author[0000-0002-3726-4881]{Zhoujian Zhang}
\affiliation{The University of Texas at Austin, Department of Astronomy, 2515 Speedway, C1400, Austin, TX 78712, USA}

\author[0000-0003-2232-7664]{Michael C. Liu}
\affiliation{Institute for Astronomy, University of Hawai'i, 2680 Woodlawn Drive, Honolulu HI, 96822, USA}

\author[0000-0003-4908-4404]{Jessy Jose}
\affiliation{Indian Institute of Science Education and Research (IISER) Tirupati, Rami Reddy Nagar, Karakambadi Road, Mangalam (P.O.), Tirupati 517 507, India}

\author{Belinda Damian}
\affiliation{CHRIST (Deemed to be University), Hosur Road, Bengaluru 560029, India}

\author[0000-0003-1634-3158]{Tanvi Sharma}
\affiliation{Institute of Astronomy, National Central University, 300, Zhongda Road, Zhongli, Taoyuan 32001, Taiwan}

\author[0000-0001-5579-5339]{Micka\"{e}l Bonnefoy}
\affiliation{Univ. Grenoble Alpes, CNRS, IPAG, F-38000 Grenoble, France}

\author{Yumiko Oasa}
\affiliation{Faculty of Education, Graduate School of Science and Engineering, Saitama University, 255 Shimo-Okubo, Sakura-ku, Saitama, 338-8570, Japan}

\begin{abstract}
We report the discovery of substellar objects in the young star cluster IC\,348 and the neighboring Barnard~5 dark cloud, both at the eastern end of the Perseus star-forming complex. The substellar candidates are selected using narrowband imaging, i.e., on and off photometric technique with a filter centered around the water absorption feature at 1.45~\micron, a technique proven to be efficient in detecting water-bearing substellar objects. Our spectroscopic observations confirm three brown dwarfs in IC\,348. In addition, the source WBIS~03492858+3258064, reported in this work, is the first confirmed brown dwarf discovered toward Barnard\,5. Together with the young stellar population selected via near- and mid-infrared colors using the Two Micron All Sky Survey and the Wide-ﬁeld Infrared Survey Explorer, we diagnose the relation between stellar versus substellar objects with the associated molecular clouds. Analyzed by Gaia EDR3 parallaxes and kinematics of the cloud members across the Perseus region, we propose the star formation scenario of the complex under influence of the nearby OB association.
\end{abstract}



\section{Introduction} \label{sec:intro}


Stars are formed in groups out of dense molecular fragments, whereas planets are condensed in young circumstellar disks.  Brown dwarfs \citep[mass $\la 0.07~M_\sun$,][]{Dupuy2017} and planet-mass objects (mass $\la 0.013~M_\sun$), collectively called substellar objects, are not massive enough to sustain core hydrogen fusion, and therefore cool and fade after birth, even though massive brown dwarfs are expected to undertake core fusion by short-supplied deuterium and lithium, or even by protons before the core becomes degenerate (e.g., \citet{Rebolo1996, Forbes2019}).
Young substellar objects are hot and bright, therefore detectable in nearby star-forming regions.
For example, a 5$~M_{\rm Jup}$ planet is expected to have $T_{\rm eff} \sim1800~K$, and $\log L/L_\sun\sim -3.5$ at age of 1~Myr, becomes $T_{\rm eff} \sim1100$~K, and $\log L/L_\sun\sim -4.4$ at age of 10~Myr, and continues to fade to $T_{\rm eff}\sim600$~K, and $\log L/L_\sun\sim -5.7$ at age of 0.1~Gyr \citep{burrows97}, corresponding to absolute $J$ magnitude, respectively, of $M_J\sim11$~mag (1~Myr), $M_J\sim14$~mag (10~Myr), and $M_J\sim17$~mag (0.1~Gyr),
respectively \citep{Allard2001,Baraffe2003}. This means such an object at an age of 1~My would have an apparent magnitude of $m_J\sim18$~mag at 200~pc given a moderate extinction $A_J=1$~mag.

A reliable catalogue of substellar members is the first crucial step to supplement the stellar population within the same cloud, i.e., to derive the mass function, and to test theories of stellar versus substellar formation. At the lowest-mass end, the free-floating planets serve as a comparison sample to exoplanets, as to the similarities and differences in formation and evolution \citep{Lucas2000}; e.g., to test the progression of cloud fragmentation to very low masses or the existence of a population of ejected giant planets \citep{Sumi2011}.

Wide-field surveys have played a major role in recognition of substellar objects in the solar neighborhood, such as by Pan-STARRS \citep[PS1;][]{Chambers2016}, the Wide-Field Infrared Explorer \citep[WISE;][]{Wright2010}, the UKIRT Infrared Deep Sky Survey \citep[UKIDSS;][]{Lawrence2007}, the Two Micron All Sky Survey \citep[2MASS;][]{Skrutskie2006}, the DEep Near-Infrared Survey of the Southern sky \citep[DENIS;][]{Epchtein1994}, the Sloan Digital Sky Survey \citep[SDSS;][]{Blanton2017}, the VISTA Hemisphere Survey \citep[VHS;][]{McMahon2019}, etc. Interstellar dust extinction makes it difficult in optical and near-infrared wavelengths to distinguish between background giant stars and equally faint substellar objects. The largest limiting factor for these surveys is the need for spectroscopic confirmation of photometrically selected candidates, which is very time-consuming unless a highly reliable candidate list is available. Supplementing the usual broadband photometry to diagnose cool atmospheres, we use a custom IR filter (with a bandpass width 6\%) centered around 1.45~\micron\ to detect the flux suppression due to H$_2$O absorption commonly seen in the spectra of low-mass stars, brown dwarfs, and planetary-mass objects \citep{Allers2020}. The combination of the water band (referred to as the $W$-band) photometry along with $J$ and $H$ magnitudes proves effective in distinguishing young substellar objects from reddened background stars that lack water absorption in their atmospheres \citep{Jose2020,Dubber2021}.

The Perseus molecular cloud complex, located to the west of the Per OB2 association, is active in star formation, with a few distinct subregions including two young populous star clusters, IC\,348 and NGC~1333, and dark clouds such as L1468, L1448, L1451, L1455, Barnard~1, and Barnard~5 (B5, also known as L1471; \citet{Jorgensen2006, Rebull2007}. Here we report identification of substellar objects in IC\,348 and in B5, both to the eastern end of the complex. This list, together with the young stellar counterparts, provides two contrasting samples, a crowded (IC\,348) versus a relatively isolated (B5) environment.

IC\,348 ($\alpha=03{^h}44{^m}30{^s}.78$, $\delta=+32$\arcdeg$10$\arcmin$19$\arcsec$.0$, J2000) harbors nearly 500 known members \citep{Luhman2016} with the most massive being a B5 star (HD\,281159). Distance determinations to the cluster range from $321\pm10$~pc by \citet{Ortiz-Leon2018} using VLBA, $320\pm26$~pc by \citet{Gaia_DR2_b, Gaia_DR2_a}, to $295\pm15$~pc by \citet{Zucker2018}. The age of the cluster has been estimated to be 2--3~Myr \citep{Luhman2003}. However, the young population seems to have a lower disk fraction given its age \citep{Cieza2015}.  Indeed, \citet{Ruiz-Rodriguez2018} using ALMA detected typical disk masses in the Class~II objects in IC\,348 to be lower than in Taurus, Chamaeleon, or $\sigma$~Ori, all with ages 1--5~Myr, thus indicating an older age (6~Myr) for IC\,348 \citep{Bell2013, Cottaar2014}. In a separate site $\sim10\arcmin$ to the southwest of the star cluster, ongoing star formation (1--2~Myr) is witnessed to be spatially coincident with dense clouds, protostars, and Herbig-Haro objects, suggestive of noncoeval episodes of star birth in the region \citep{Tafalla2006}.

Farther to the northeast of IC\,348 ($\sim3$~pc), at the far end of the cloud complex, is B5, an isolated globule with cold \citep{beichman1984} and large dust grains \citep{Bhatt1986}. More distant than IC\,348 \citep[350~pc;][]{herbig1983, langer1989}, and spanning some $45\arcmin$ in the sky, B5 is relative massive ($\sim10^{3}$~M$_\sun$, \citet{goldsmith1986, langer1989}) and known also to have ongoing star formation, yet hosting several dense cores \citep{fuller1991, Olmi2005, Hatchell2005, Kirk2006}, fragmenting cloud filaments \citep{Pineda2015}, and Herbig-Haro objects \citep{Walawender2005a}. An overall low star formation efficiency is in vast contrast to the adjacent IC\,348 cluster which is prolific in star birth \citep{Hatchell2007}.

This paper is a continuation of the W-band series on substellar objects using the novel $W$-band technique in nearby star-forming regions (\citet{Allers2020, Jose2020, Dubber2021}, Dubber et al. (2022) submitted). We report here deep imaging surveys using the $W$-band, combined with $J$ and $H$ bands, to identify young brown dwarfs in IC\,348 and B5. While the young stellar population in IC\,348 is relatively well known, our focus in this cloud is on newly found brown dwarfs. For B5, on the other hand, in addition to the substellar candidates found with the $W$-band imaging, we have conducted a census of young stars with near- and mid-infrared colors. Finally we have characterized the distance and kinematics of the young objects found in IC\,348 and in B5, and will discuss them in the context of the overall morphology and motion of the whole Perseus cloud complex as evidence of the star formation history therein.  

\section{Observation and Data Analysis}
\label{sec:observation_data}

The data used in this study include the water-band images to select substellar candidates, plus archival near- and mid-infrared photometry from 2MASS and WISE to identify young stars with dusty circumstellar disks, and Gaia EDR3 for parallax and kinematics.  

\begin{figure*}
\centering
\includegraphics[width=1.20\textwidth]{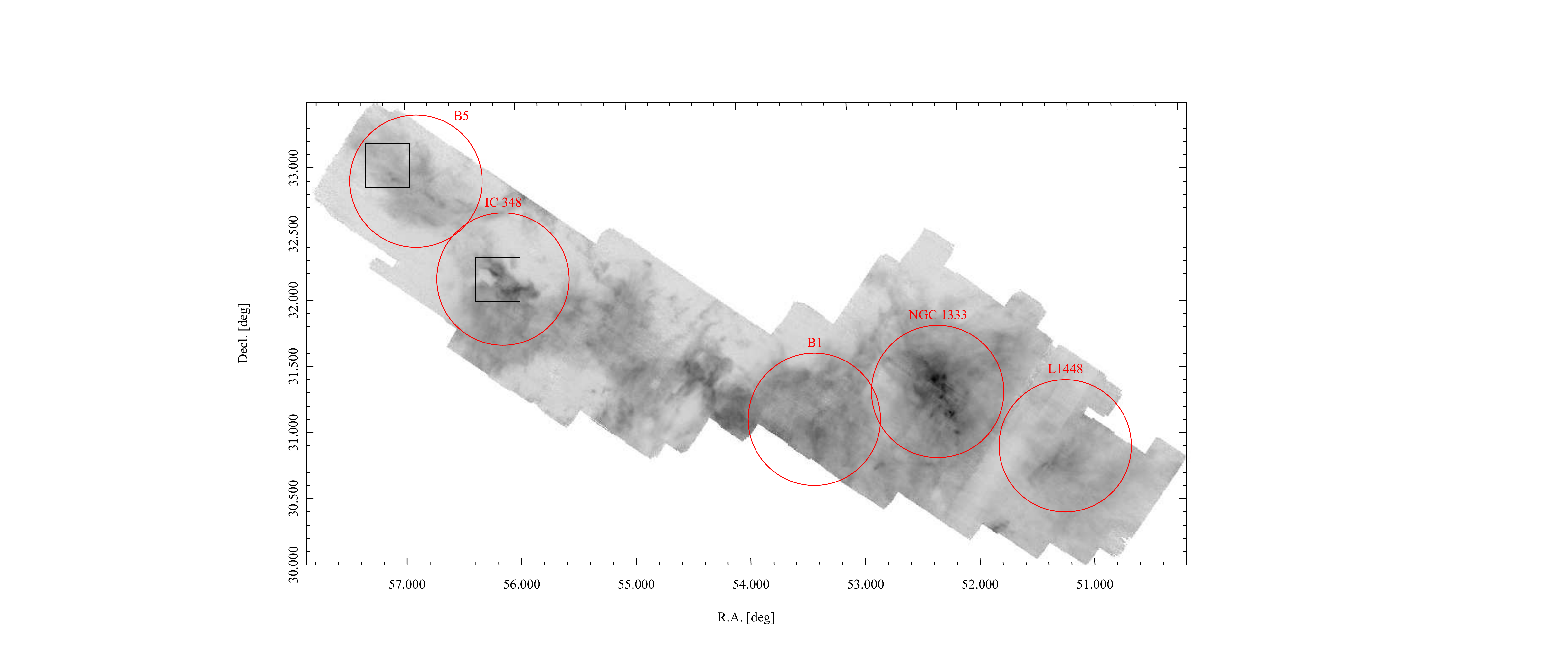}
\caption{\textsuperscript{12}CO (J=$1-0$) FCRAO data of Perseus molecular cloud \citep{Ridge2006b}. The black squares show the fields toward IC\,348 and B5 covered by the W-band imaging survey. The red circles mark the fields of $\sim30\arcmin$ radius analyzed by using Gaia EDR3 data, including three other regions, Barnard~1, NGC~1333, and L1448 (see Sec.~\ref{sec:Ind_yso_b5} for further details).}
\label{fig:B5_IC348_CO}
\end{figure*}

\subsection{CFHT/WIRCam Water-Band Imaging}

IC\,348 was observed on March 18\textsuperscript{th}, 2016, July 16\textsuperscript{th}, 2016, and December 08\textsuperscript{th}, 2017; and B5 was observed on October 23\textsuperscript{rd} 2015 (all in UT), with the Canada-France-Hawaii Telescope (CFHT) using the Wide Field Infrared Camera \citep[WIRCam;][]{Puget2004}. With a field of view of $\sim 20\arcmin \times 20\arcmin$ and $\sim 0.306\arcsec$ pixels, we used a 21-point dithering pattern to fill the gaps between the four detectors of WIRCam and to subtract the sky background. Images were processed by the CFHT `I`iwi pipeline. Darks exposure times were 10 sec in $J$-, and 65 sec for $H$- and $W$- bands. Normalized flat-field corrections for each filter was constructed after each observing run. Before stacking the images, saturated and bad-pixel were flagged, non-linearity were corrected, and images were dark subtracted, flat-field corrected, sky-subtracted, cross-talk removed and 1/f noise corrected by `I`iwi. The total integration time for $J$, $H$, and $W$~bands for IC\,348 was 170, 210, and 1495~s, respectively, and for B5 it was 50~s ($J$), 75~s ($H$), and 455~s ($W$), respectively.

To calibrate the $J$-band and $H$-band stacks, we adjusted the zero point by including a color term between 2MASS and the MKO system of WIRCam by clipping the magnitudes in the range $14.5\leq J \leq 16$~mag and $14.5\leq H \leq 15.5$~mag, with the lower limit being the saturation limit of the WIRCam, and the upper limit being decided by the errors of 2MASS magnitudes (i.e., $< 0.1$~mag). We used the color correction from \citet{Leggett2006} to convert the 2MASS $J$ and $H$ magnitudes to the MKO system, and then found the zero point offset for each of the $J$ and $H$ bands. 

In absence of 2MASS equivalents, the zero point of the $W$-band photometry was determined with a different strategy. Following \cite{Allers2020}, we adopt a $Q$ index as a measure of the flux suppression in the $W$-band relative to the dereddened continuum between $J$- and $H$-bands, as shown in Eq.~\ref{eq_Q}, 

\begin{equation}
    Q = (J - W) + e (H - W),
\label{eq_Q}
\end{equation}
where $J$ and $H$ are the magnitudes already calibrated, $W$ is the $W$-band magnitude to be calibrated, and $e = (A_J - A_W) / (A_W - A_H)$ is an empirical coefficient to account for the reddening across the bands.  To calibrate the $W$ magnitude, the median $Q$ value, $Q_\textrm{f}$, for the majority of bright stars most likely lacking water absorption is adjusted so that $Q_{\rm f} \approx 0$; that is, 

\begin{equation}
  W = (e H + J - Q_\textrm{f})/(1.0 + e), 
\label{eq_W}
\end{equation}
which, based on our synthetic photometry for a field star (i.e., M0) yielding $Q_\textrm{f} = -0.043$, and have adopted $e=1.61$ \citep[][]{Allers2020}. The $W$-band zero point is then applied to offset the WIRCam instrumental magnitude. The calibrated $J$, $H$, and $W$ magnitudes together allow the $Q$ index to be used to quantify the flux suppression in the $W$ band to select water-bearing candidates, with the methodology to be detailed later in Sec.~\ref{sec:qvalue}

\subsection{Spectroscopic Observations}

Spectroscopic follow-up observations were conducted with the SpeX spectrograph on the IRTF \citep{Rayner2003}. The targets were chosen as being sufficiently bright ($H < 17$~mag), and were checked whether they were already known brown dwarfs \citep{Muench2007, Luhman2016, AlvesdeOliveira2013}. We incorporated available information from SIMBAD, 2MASS, PanSTARRS, and UKIDSS photometry. In order to select targets for spectroscopic follow-up with no photometric peculiarity.

\begin{figure}[htbp!]
\centering
  \includegraphics[width=\columnwidth]{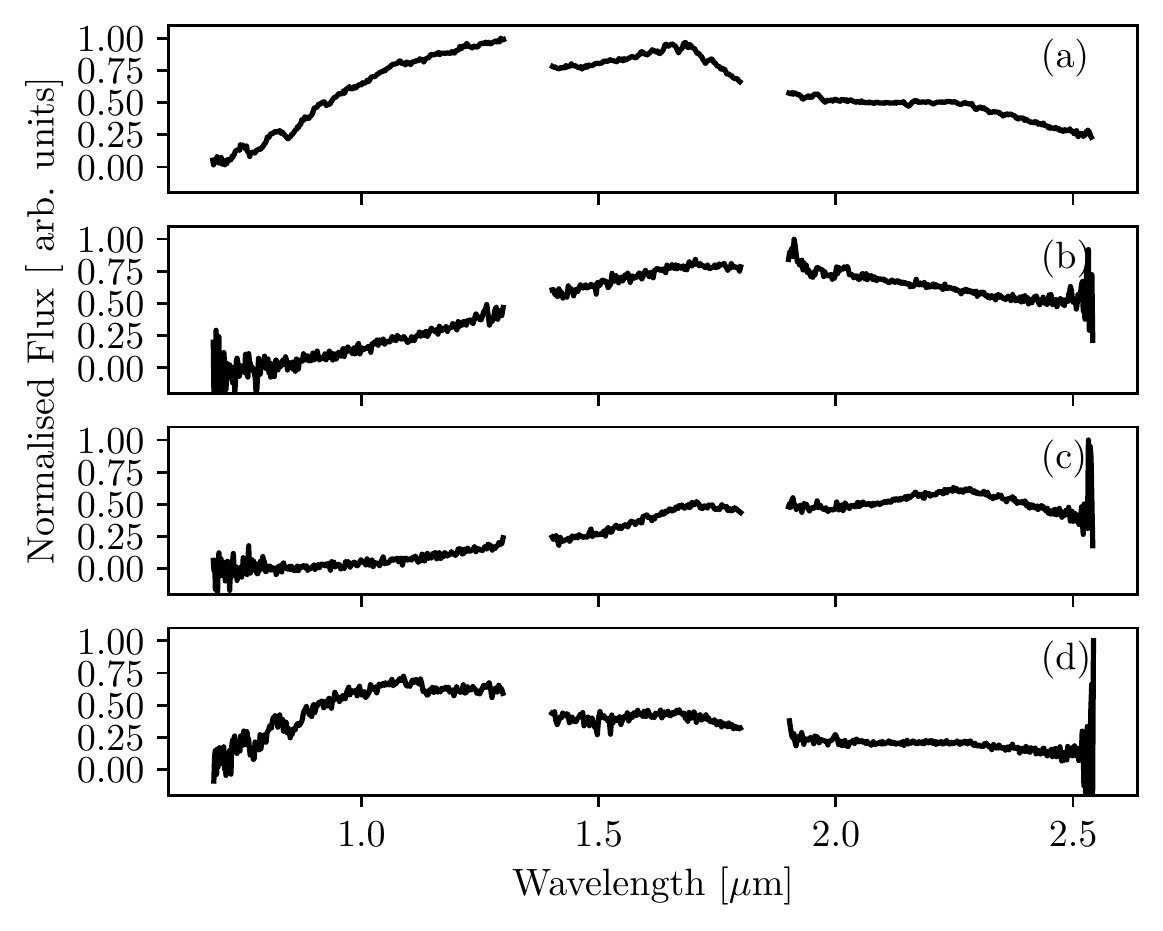}
  \caption{The observed flux calibrated and normalised IRTF/SpeX spectra of the targets in IC\,348: (a) 03434538+3201041, (b) 03435016+3204073, (c) 03441864+3218204, and (d) 3441864+3218204
          }
  \label{fig:observed_spectra_ic348}
\end{figure}

Three targets in IC\,348 were observed on February 18\textsuperscript{th}, 2017, and one target on January 12\textsuperscript{th}, 2022 (all in UT). A standard ABBA nodding pattern was utilized to obtain the sky and target data, with a total integration time of 1473~s for each target. The observation were taken in the prism mode using a $0.8\arcsec$ slit, which corresponds to an average resolving power of $R\sim100$. Flat field frames and wavelength calibration using an argon lamp were taken before or after a set of observations. We used SPEXTOOL \citep[V4.1][]{Cushing2005}, an IDL-based data reduction package, to extract the spectra. An A0 standard star, either HD\,25152 or HD\,23512, was observed either before or after a science target for telluric corrections. The IRTF spectra for IC\,348 are shown in Fig.~\ref{fig:observed_spectra_ic348}.

\begin{figure}[tbh!]
  \includegraphics[width=\columnwidth]{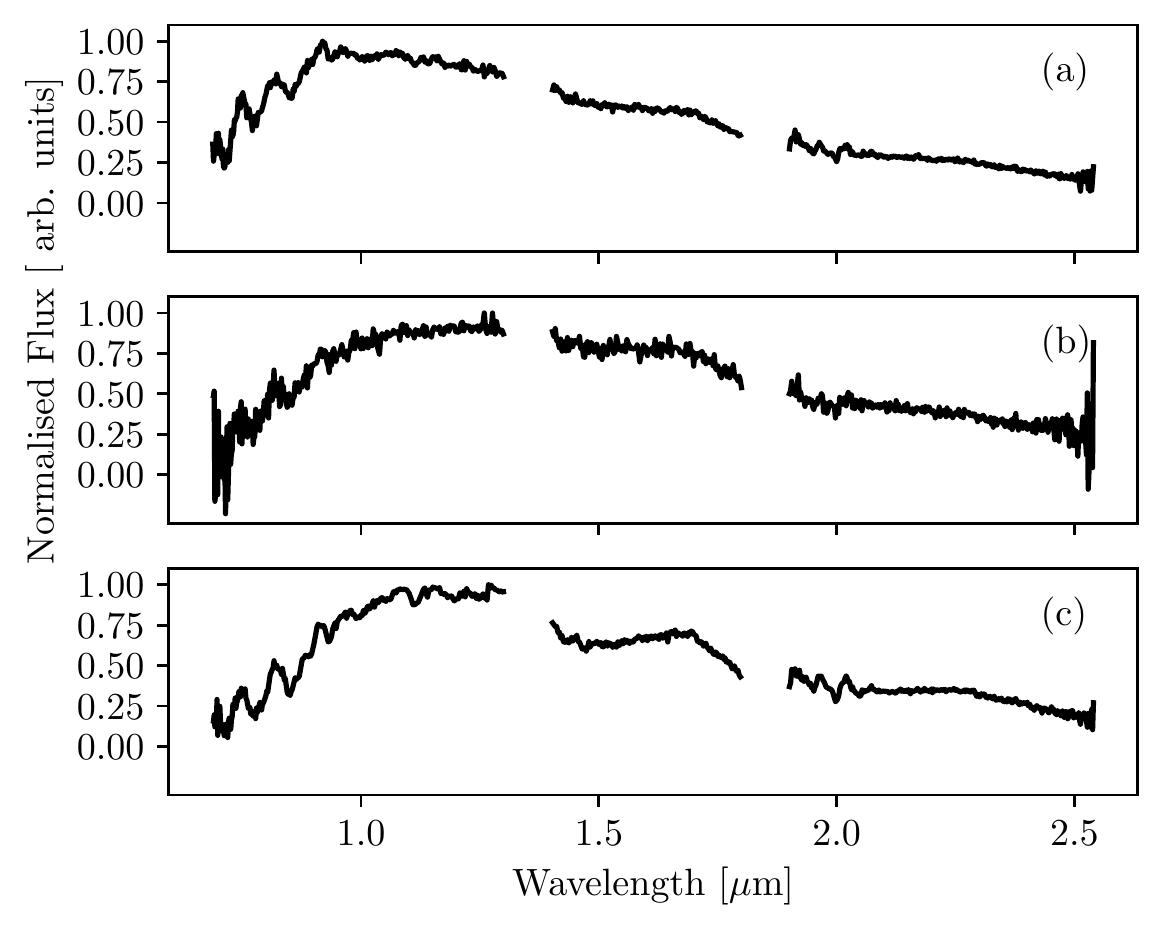}
  \caption{The observed flux calibrated and normalised IRTF/SpeX spectra of the targets in Barnard~5: (a) 0348154+330918, (b) 0348362+325905, and (c) 0349286+325806.
           }
  \label{fig:observed_spectra_b5}
\end{figure}

Three B5 targets were observed on January 12\textsuperscript{th}, 2022 (UT), also in the prism mode, but with a $0.5\arcsec$ slit and with the standard star HD\,22859, following the same data reduction procedure. The spectra of the three B5 targets are shown in Fig.~\ref{fig:observed_spectra_b5}. 

\subsection{WISE and 2MASS}

The census of the young stellar population in IC\,348 has been relatively comprehensive in the literature, therefore our attention was mainly on the substellar objects, but for B5, we pay attention to the stellar as well as substellar objects in general.  We conducted a search within a field of $\sim30\arcmin$ toward B5 for young stellar candidates using near- and mid-infrared colors, with 2MASS \citep{cutri2003} photometry in $J$ (1.25~\micron), $H$ (1.65~\micron), and $K_s$ (2.15~\micron) bands, and WISE \citep{Cutri2012} in $W1$ (3.4~\micron), $W2$ (4.6~\micron), $W3$ (12~\micron), and $W4$ (22~\micron) bands. There are 10588 WISE sources within the B5 field, of which 1140 have photometric quality flag `AAA' (206) or `AAB' (934), i.e., with an uncertainty less than 0.1~mag (for A) or 0.3~mag (for B) in respective bands, whereas among the 4282 2MASS sources, 1985 have photometric quality flag `AAA' with uncertainty $\la 0.1$~mag.

\subsection{Gaia EDR3}

We made use of the Gaia Early Data Release~3 (Gaia EDR3) \citep{GaiaCollaboration2021} which provides homogeneous astrometry (coordinates, trigonometric parallax, and proper-motion) of about 1.5 billion sources of the entire sky. Broad-band photometry in $G, G_{\rm BP},$ and $G_{\rm RP}$ with passbands of 330--1050~nm, 330--680~nm, and 630--1050~nm respectively are also available. We analyzed the Gaia EDR3 data towards IC\,348, and B5, plus three other regions in the Perseus complex to diagnose the parallax and kinematic properties.  


\section{Young Stellar and Substellar Members}

\subsection{Selection of Brown Dwarf Candidates}\label{sec:qvalue}

\begin{figure}[htbp!]
\centering
  \includegraphics[width=\columnwidth]{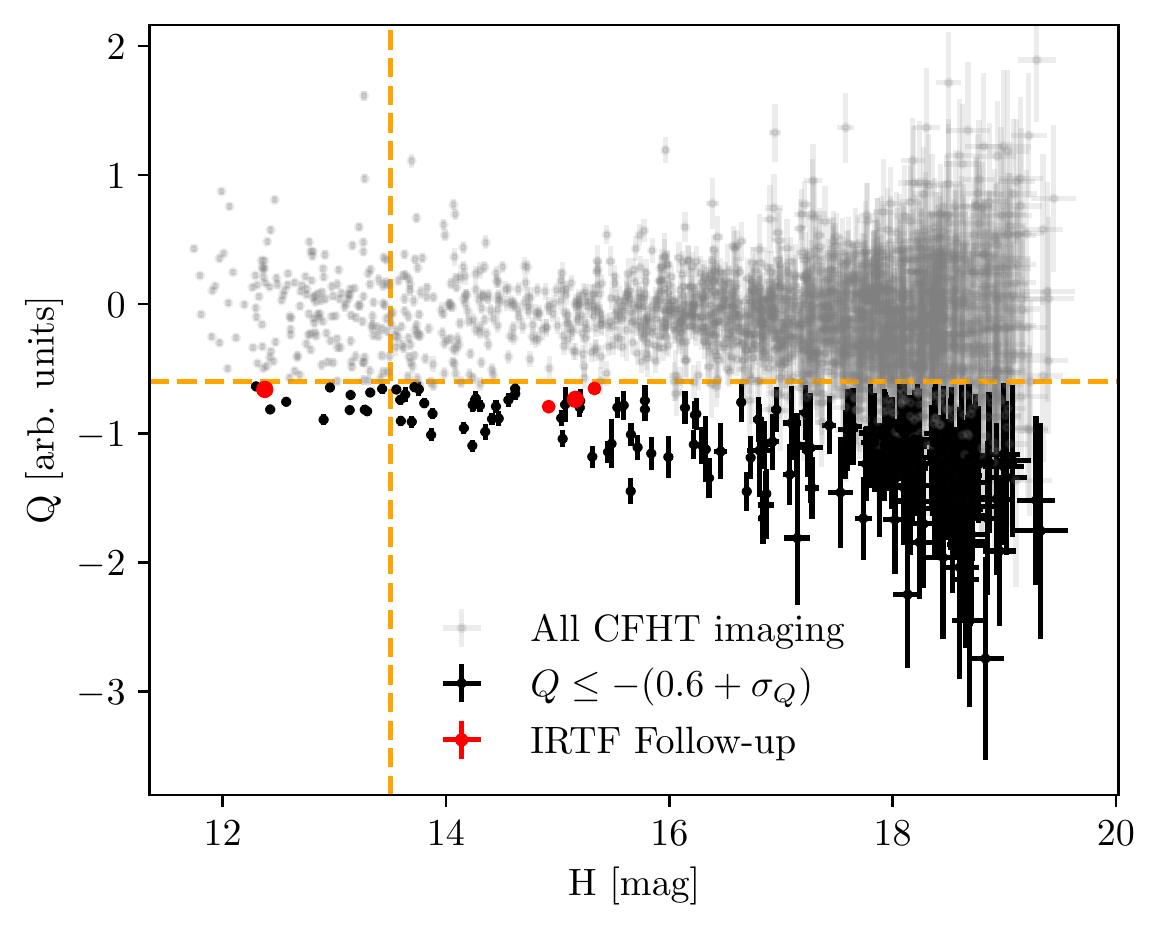}
  \caption{The $Q$ index versus $H$ magnitude in IC\,348. Of the total 1943 sources (in gray), those likely to exhibit flux suppression due to water absorption, i.e., with $Q + \sigma_Q \leq -0.6$ are marked in black (163 targets) with error bars. Those selected for spectroscopic follow-up are represented in red. The orange dashed line of $H=13.5$~mag indicates the nominal photometric saturation limit, whereas the one with $Q = -0.6$ denotes the substellar boundary for young objects. Note that the targets selected for spectroscopic follow-up satisfy the water absorption and cool temperature criteria i.e., ($W1-W2$)$>0.10$ and ($z_{PS1}-J_{CFHT}$)$\leq 2$}
  \label{fig:h_vs_q_ic348}
\end{figure}

\begin{figure}[htbp!]
\centering
  \includegraphics[width=\columnwidth]{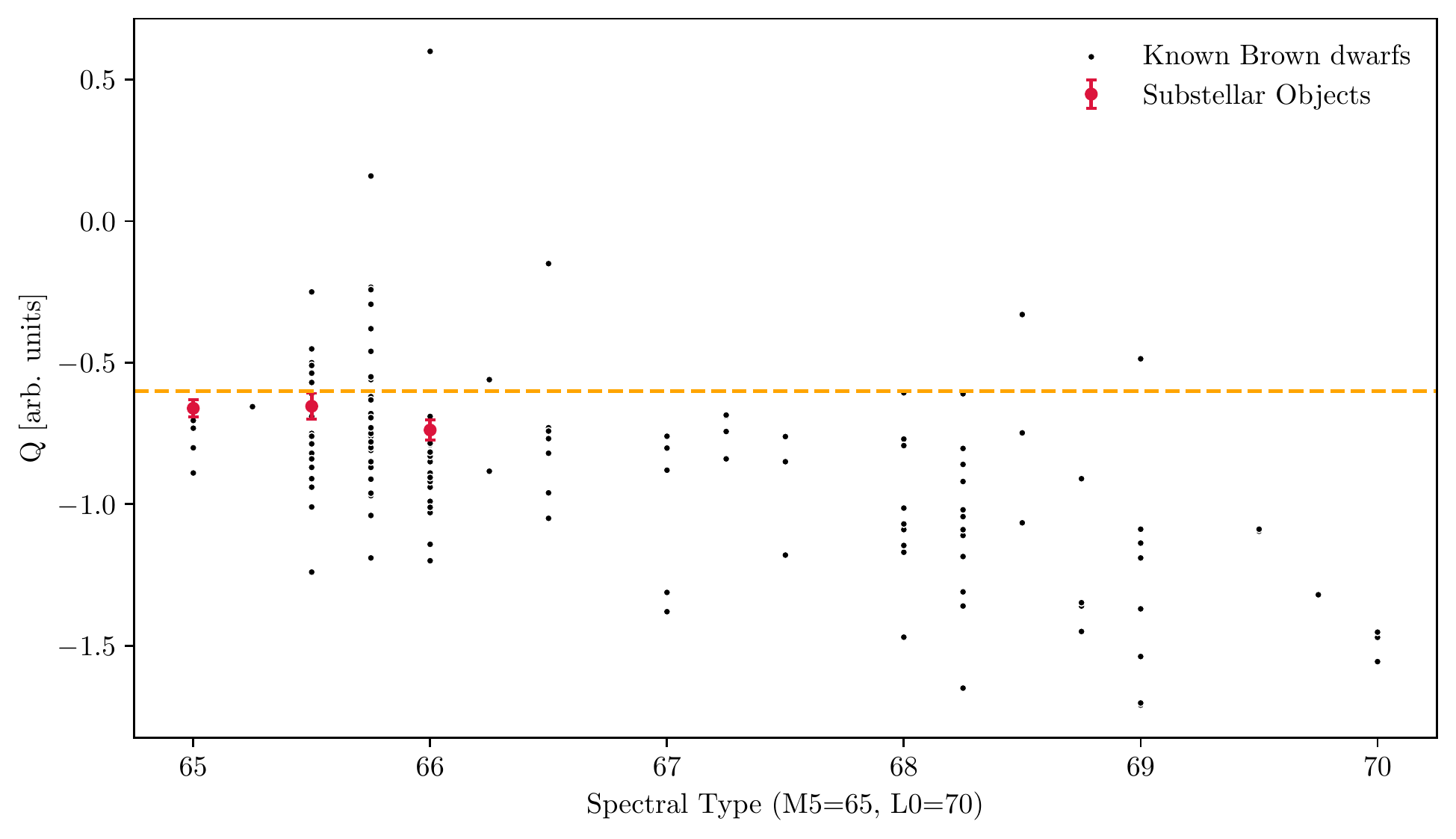}
  \caption{Spectral type versus $Q$ index of all known brown dwarfs in IC\,348. The known brown dwarf list is taken from, \citet{Muench2007}, \citet{Luhman2016}, and \citet{AlvesdeOliveira2013}. New discoveries from this work (confirmed spectroscopically) i.e., substellar objects are shown in red with respective error bars. The orange dashed line denotes $Q = -0.6$, below which the $Q$ value indicates a spectral type later than M6, corresponding to a substellar boundary for young objects.}
  \label{fig:known_bds_ic348}
\end{figure}

\begin{figure}[htbp!]
 \centering
  \includegraphics[width=\columnwidth]{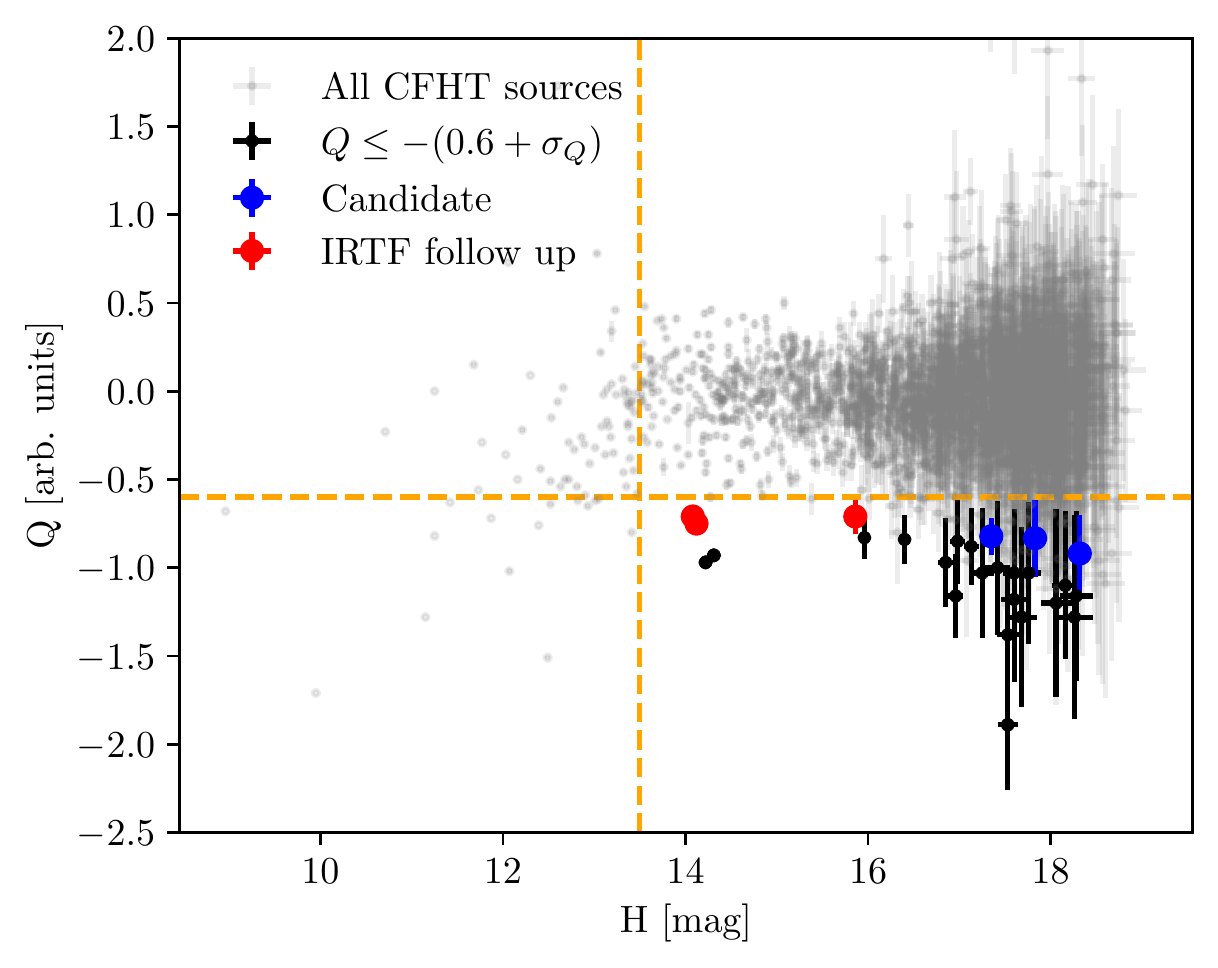}
  \caption{The $Q$ index versus $H$ as in Figure.~\ref{fig:h_vs_q_ic348} but for Barnard~5 (here with 2089 sources in gray). The black symbols mark those satisfying criteria of $Q \leq -(0.6 + \sigma_Q )$ and $H > 13.5$~mag (23 targets). The red symbols mark candidates that have been followed up spectroscopically with IRTF/SpeX, and the blue symbols show candidates too faint to be observed with IRTF/SpeX. Note that both the red and blue symbol targets satisfy the water absorption and cool temperature criteria i.e., ($W1-W2$)$>0.10$ and ($z_{PS1}-J_{CFHT}$)$\leq 2$.}
  \label{fig:h_vs_q_b5}
\end{figure}

The use of the $Q$ index to distinguish brown dwarfs from reddened background objects \citep[see][for detailed descriptions]{Allers2020, Jose2020, Dubber2021} is much more effective than using broad-band photometric colors alone. With the definition of $Q$ in Eq.~\ref{eq_Q}, and the ratio of extinction in the three bands, which depends on the adopted reddening law. To set this value, we calculated synthetic photometry for an M0 spectra taken from \citet{Kirkpatrick2010}, reddened by $A_V=10$~mag and $R_V=3.1$ reddening law of \citet{Fitzpatrick1999}. Next, we compared the reddened and unreddened synthetic photometry of the M0 spectra and adopted extinction coefficient to be $e=1.61$  (see \citet{Allers2020}). A typical contaminant star would have $Q \sim 0$ (without water absorption), whereas for a later-type object the $Q$-index would be negative due to water absorption, as presented in Figure~\ref{fig:h_vs_q_ic348} for IC\,348. By using the synthetic photometry in $J$, $H$, and $W$ of young objects \citep{Muench2007, Allers2013} and field dwarfs \citep{Cushing2005}, we concluded that objects with spectral types later than M6 would have $Q < -0.6$. Including the uncertainty $\sigma_{Q}$, estimated by calculating the errors in quadrature, namely, 

\begin{equation}
    \sigma_Q = \sqrt{\sigma_J^2 + (e\, \sigma_H)^2 + [(1 + e) \, \sigma_W]^2}, 
\end{equation}
we therefore considered sources with $Q \leq -(0.6 + \sigma_{Q})$ for plausible presence of water absorption, i.e., as substellar candidates, and with $H >14$~mag to avoid photometric saturation. The limitation to our approach is that, if a young substellar object is still embedded in the parental material and/or have a substantial mass accretion, thereby entailing strong veiling \citep{Luhman1999}, our technique may not be able to identify it.
In addition to molecular features, criteria of cool atmospheres can be added \citep[e.g.,][]{chiang2015}. For this we imposed the condition $(W1-W2)$ $>0.10$, the intrinsic color for a spectral type later than M6 \citep{Kirkpatrick2011}. We also cross-matched with Pan-STARRS~1 \citep{Chambers2016} to remove objects with bright optical counterparts, i.e., ($z_{PS1} - J_{CFHT}$) $\leq 2$ \citep{Knapp2004}. Because of the large WISE pixel scale, we inspected visually are the Pan-STARRS~1 and CFHT images for each candidate to ensure the ($W1-W2$) color not to be affected by blends with other sources. Combining both the water absorption and cool temperature criteria i.e., ($W1-W2$)$>0.10$ and ($z_{PS1}-J_{CFHT}$)$\leq 2$, and by excluding any known members found in literature (i.e., \citet{Muench2007, Luhman2016, AlvesdeOliveira2013}), obtained a final list of four targets in IC\,348 and three targets in B5, shown in Fig.~\ref{fig:h_vs_q_ic348}, and Fig.~\ref{fig:h_vs_q_b5}, for follow-up spectroscopy.
In Figure~\ref{fig:known_bds_ic348} we demonstrate that an increasingly negative value of $Q$ corresponds to an object with a later type. Here we show $Q$ versus spectral type (SpT) of all the known members in the literature and spectroscopically identified substellar objects in IC\,348.

\subsubsection{Spectral Fitting}

We compared our spectra with library templates for spectral typing. Those for field brown dwarfs are taken from the SpeX library \citep{Cushing2005, Rayner2009} whereas spectra for young brown dwarfs are adopted from \citet{Luhman2017}. Each standard star belongs to one of the three age groups: VL-G (very low gravity $\approx 1$~Myr), INT-G (intermediate gravity, with two subgroups: I for 10~Myr and Y for 5~Myr), or FLD-G (gravity for much older field stars). The data are dereddened by $A_{V}$ ranging from 0 to 40~mag, using the reddening law from \citet{Fitzpatrick1999}. For further details on spectral classification, please see \citet{Dubber2021}. 

\begin{figure*}
\gridline{\fig{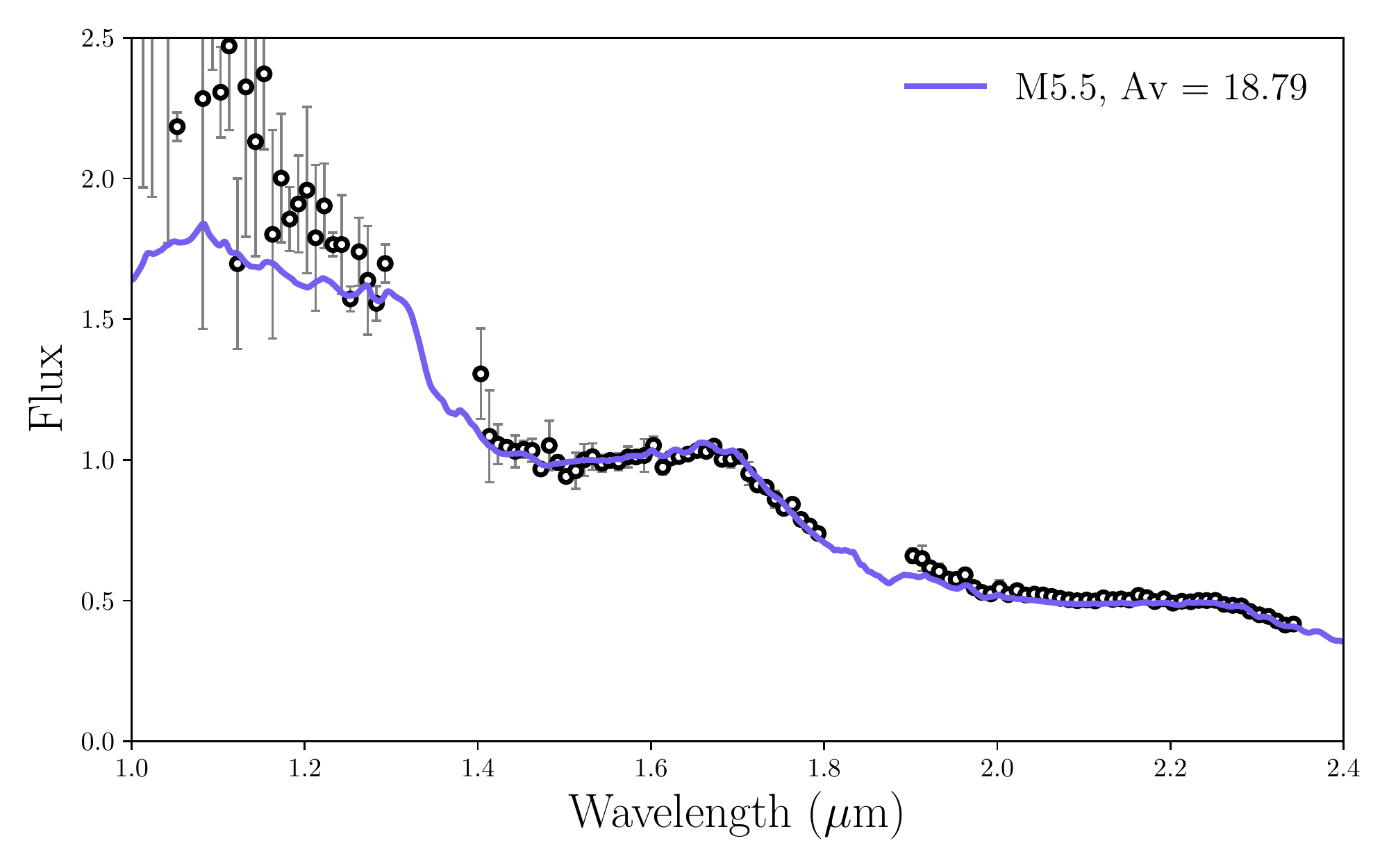}{0.5\textwidth}{(a)~034350+320407}
          \fig{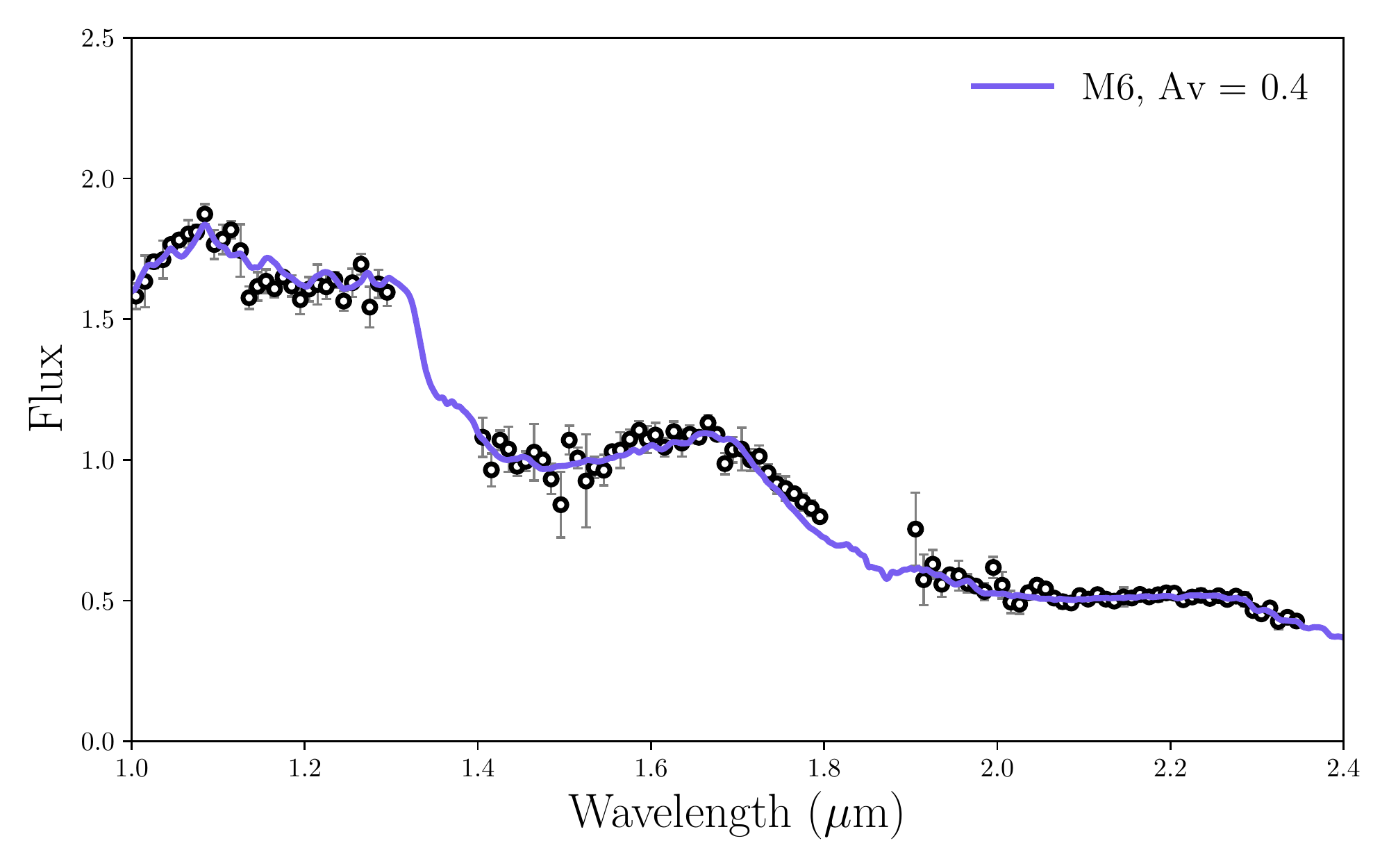}{0.5\textwidth}{(b)~034418+321820}
          }
\gridline{\fig{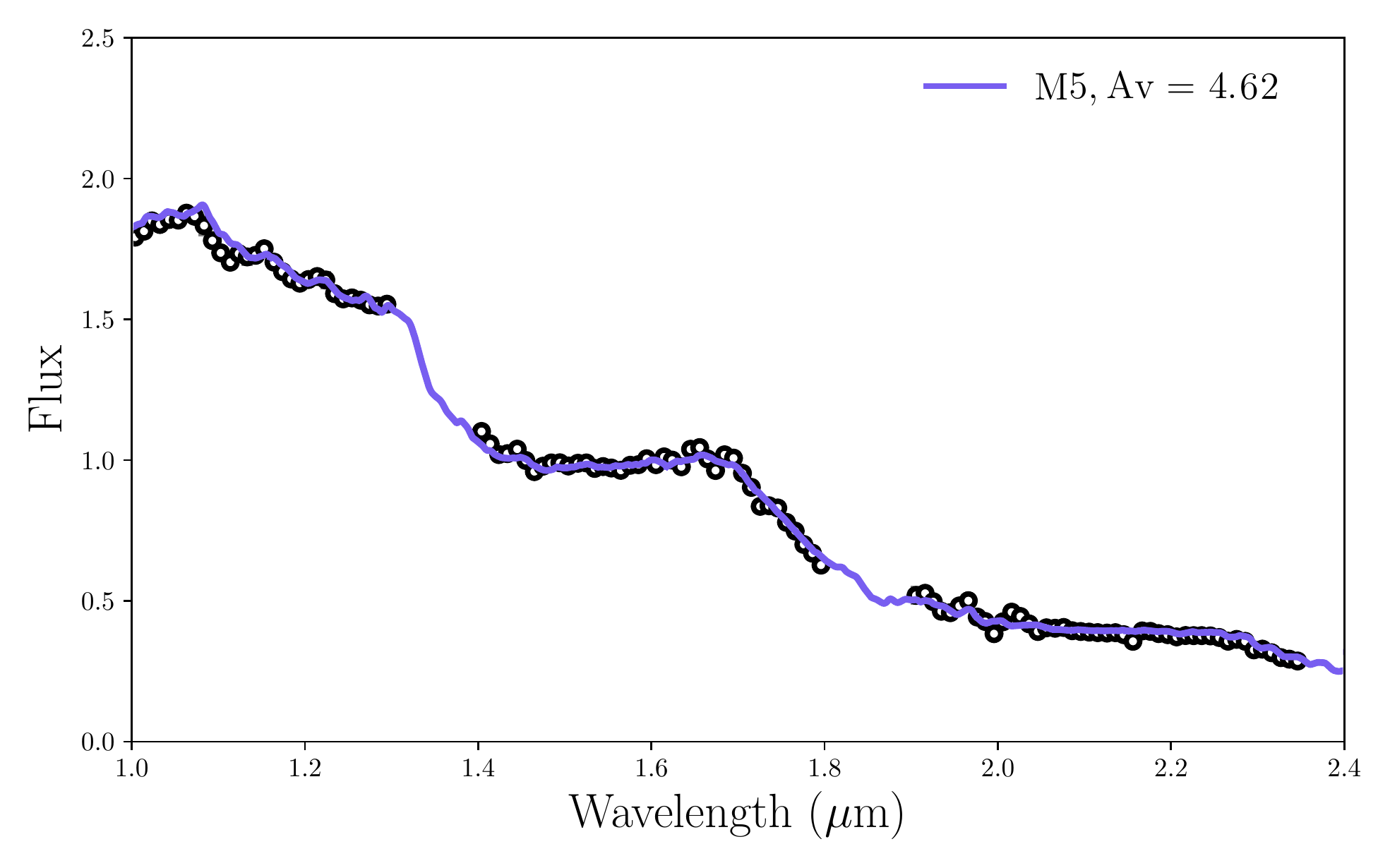}{0.5\textwidth}{(c)~034515+321821}
          \fig{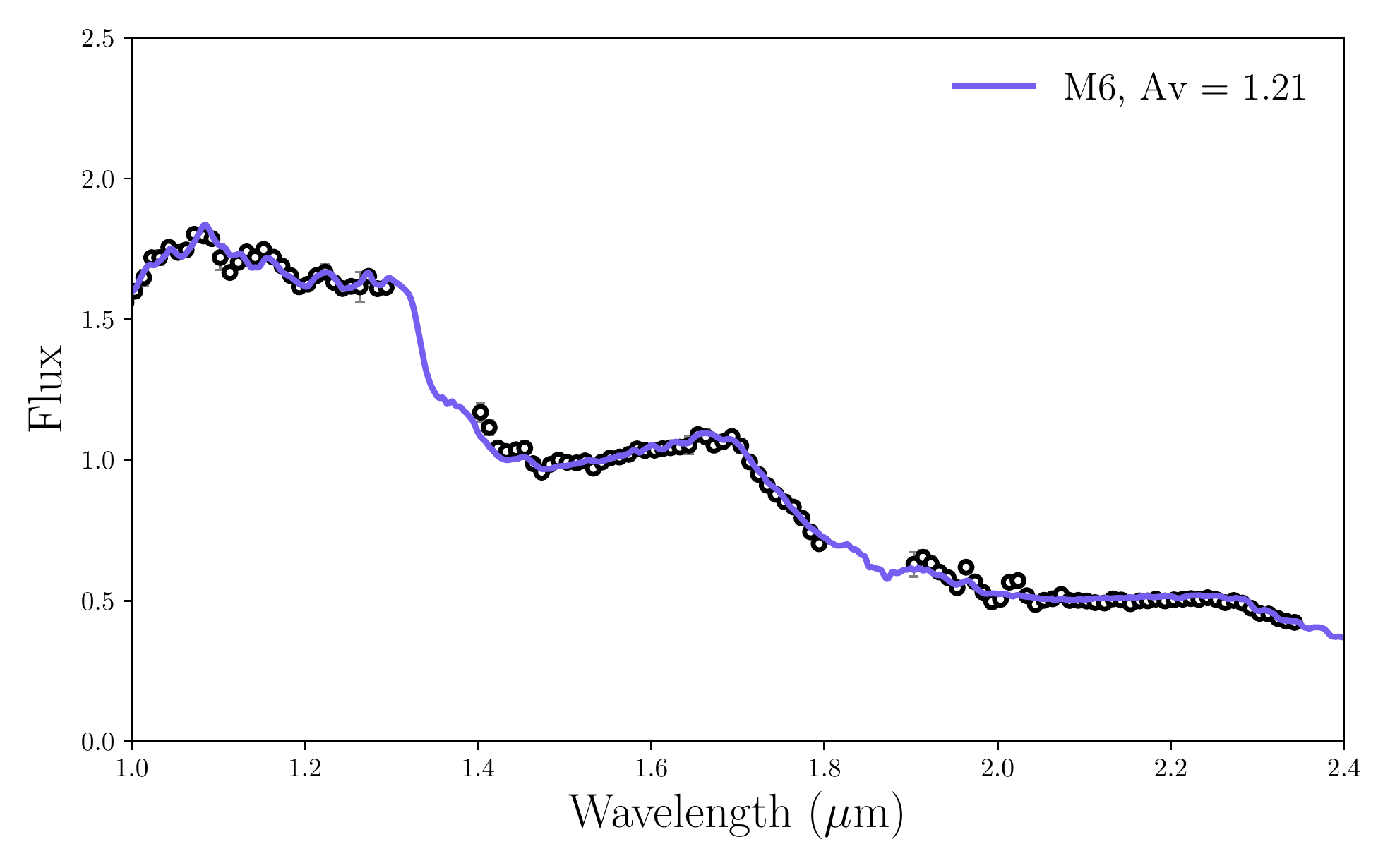}{0.5\textwidth}{(d)~034928+325806}
         }
\caption{The dereddened spectra (in black) of the three candidates in IC\,348 are shown in (a), (b) and (c) and one in B5 is shown in (d), with the best-fit template (in purple) shown in each case.  
        }
\label{fig:spectral_fit}
\end{figure*}

Spectra of young objects later than M5 show a characteristic inverted ``V'' (triangular) shape in the $H$~band \citep{Cushing2005}. This feature, where $H$ band is dominated by the strong water absorption bands peaking around 1.70~\micron, was first reported by \citet{Lucas2001} seen in young cool objects. The triangular peak in the $H$-band is attributed to the effect of low-pressure and low-gravity atmospheres \citep{Lucas2000, Allers2013, Allers2007}. We found three new substellar objects in IC\,348 with spectral type $\geq$M5 placing them at the boundary of the substellar regime (see Figure~\ref{fig:spectral_fit}(a), (b) and (c)).

In B5, the source 03481542+3309185 has a spectral type of M4$\pm0.5$ and can be fitted by either a field or a young template and finally assigned the object as a field dwarf. Spectral classification according to \citet{Zhang2018} this object has a spectral type of M5.2$\pm0.9$ and a low surface gravity object (VL-G). This source therefore could either be a field or a young substellar object. The source 03492858+3258064, is the first confirmed brown dwarf toward B5 (see Figure~\ref{fig:spectral_fit}(d)). Three targets were too faint to be observed with IRTF/SpeX, so we refer as brown dwarf candidates (see Table~\ref{tab:region_cc}).
The newly identified substellar objects in IC\,348 and in B5 are summarized in Table~\ref{tab:regions}. Column~1 gives the ID in the water-band survey, followed by columns 2 and 3 of the coordinates. In Columns 4 and 5 we list the $J$, and $H$ band magnitudes measured by the CFHT/WIRCam. Column 6 gives the WISE $W1$ and $W2$-bands. Column 7 gives the difference between $z$-band in PanSTARRS and $J_\mathrm{CFHT}$. The $W$ band data along with $J$, and $H$ band magnitudes, lead to the derived $Q$ parameter in Column~8 (see Eq.~\ref{eq_Q}), used to identify brown dwarf candidates. Columns 9, 10 gives the spectral type and $A_V$ obtained from spectral fitting. In Column 11 the Age estimation is based on the spectral fitting mentioned above. The bolometric correction needed to calculate $L_{\rm bol}$ in Columns 12 and $T_{\rm eff}$ in Column 13 are taken from \citet{Filippazzo2015} (for spectral type M7 or later) or \citet{Herczeg2014} (earlier than M7), depending upon the spectral type of the object.

\movetabledown=6cm
\begin{rotatetable*}
\begin{deluxetable*}{ccccclccccccc}
\tablecaption{\label{tab:regions} Properties of newly identified \textbf{substellar objects} in IC\,348 and Barnard~5.}
\tabletypesize{\scriptsize}
\tablehead{\colhead{ID} & \colhead{$\alpha$ (J2000)} & \colhead{$\delta$ (J2000)} & \colhead{$J$} & \colhead{$H$} & \colhead{$W1-W2$} & \colhead{$z_\mathrm{PS1}-J_\mathrm{CFHT}$}& \colhead{$Q$} & \colhead{$SpT$\textsuperscript{a}} & \colhead{$A_V$} & \colhead{Age\textsuperscript{b}} & \colhead{log($L_{\rm bol}/L_\sun$)} & \colhead{$T_{\rm eff}$} \\
\colhead{WBIS}&\colhead{(deg)} & \colhead{(deg)} & \colhead{(mag)} & \colhead{(mag)} & \colhead{(mag)} & \colhead{(mag)} & \colhead{} & \colhead{} & \colhead{(mag)} & \colhead{} & \colhead{} & \colhead{($10^3$~K)}}
\startdata
\multicolumn{10}{c}{IC\,348} \\
03435016+3204073$^{\dagger}$ & 55.95905 & +32.06875 & $17.64\pm 0.03$ & $15.33\pm 0.01$ & 0.59 &\nodata & $-0.65\pm 0.05$ & M5.5  & 18.79 & VL-G & $-0.65^{-0.129}_{-0.135}$ & $2.92\pm22$ \\
03441864+3218204 & 56.07767 & +32.30568 & $15.79\pm 0.01$ & $15.16\pm 0.01$ & 0.19 &1.99 & $-0.74\pm 0.04$ & M6  & 0.40 & INT-G (I) & $-2.25^{-0.094}_{-0.117}$ & $2.86\pm38$ \\
03451508+3218215 & 56.31284 & +32.30598 & $13.32\pm 0.01$ & $12.38\pm 0.01$ &0.28 & 3.04 & $-0.66\pm 0.03$ & M5 & 4.62 & INT-G (I) & $-1.54^{0.089}_{0.097}$ & $3.09\pm49$ \\ \hline
\multicolumn{10}{c}{Barnard\,5} \\
03492858+3258064 & 57.36907 & +32.96845 & $14.77\pm 0.01$ & $14.12\pm 0.01$ &0.29 &2.29 & $-0.75\pm 0.03$ & M6 & 1.21 & INT-G (I) & $-1.95^{0.076}_{0.111}$ & $2.86\pm9$ \\
\enddata
\tablecomments{\begin{itemize}
 \item[a]The uncertainty in the spectral classification type is 0.5 subtype.
\item[b]The age is estimated based on spectral fitting in \citet{Dubber2021} (please see Figure~\ref{fig:spectral_fit}).
\item[$\dagger$]This target does not have a counterpart in the Pan-STARRS (PS1) catalog. The PS1 images were visually inspected. The target is not visible in $g$,$r$, or $i$ band, and is faint in $z$ and $y$~bands.
\end{itemize}
}
\end{deluxetable*}
\end{rotatetable*}

\movetabledown=6cm
\begin{rotatetable*}
\begin{deluxetable*}{ccccclccccccc}
\tablecaption{\label{tab:region_cc} Contaminants with spectroscopic follow-up and the remaining candidate \textbf{substellar objects} in IC\,348 and Barnard~5.}
\tablehead{\colhead{ID} & \colhead{RA (J2000)} & \colhead{DEC (J2000)} & \colhead{$J$} & \colhead{$H$} & \colhead{$W1-W2$} & \colhead{$z_\mathrm{PS1}-J_\mathrm{CFHT}$} & \colhead{$Q$} & \colhead{SpT\textsuperscript{a}} & \colhead{$A_V$} & \colhead{Age} & \colhead{log($L_{\rm bol}/L_\sun$)} & \colhead{$T_{\rm eff}$} \\
\colhead{WBIS}&\colhead{(deg)} & \colhead{(deg)} & \colhead{(mag)} & \colhead{(mag)}& \colhead{(mag)} & \colhead{(mag)} & \colhead{} & \colhead{}  & \colhead{(mag)} & \colhead{} & \colhead{} & \colhead{($10^3$~K)}}
\startdata
\multicolumn{13}{c}{IC\,348} \\
03434538+3201041 & 55.93912 & +32.01781 & $16.69\pm 0.02$ & $14.92\pm 0.01$ & 0.315 &3.52 & $-0.79\pm 0.04$ & $<$M4 & 13.07 & \nodata  & $-0.91^{-0.129}_{-0.156}$ & $3.81\pm83$ \\ \hline
\multicolumn{10}{c}{Barnard\,5} \\
03481542+3309185 & 57.06426 & +33.15515 & $14.83\pm 0.01$ & $14.08\pm 0.01$ & 0.19 & 1.68 & $-0.71\pm 0.02$ & M4 & 0.85 & FLD-G & $-1.99^{0.079}_{0.108}$ & $3.09\pm53$ \\
03483616+3259054 & 57.15066 & +32.98484 & $16.78\pm 0.04$ & $15.86\pm 0.03$ & 0.10 & 2.45 & $-0.71\pm 0.10$ & $<$M4 & 2.56 & INT-G (Y) & $-1.73^{0.077}_{0.112}$ & $3.30\pm125$ \\
03482654+3252117 & 57.11059 & +32.86992 & $19.31\pm 0.08$ & $18.32\pm 0.06$ & 0.11 & 2.83 & $-0.92\pm 0.22$ & BDc & \nodata & \nodata & \nodata \\
03475398+3253030 & 56.97494 & +32.88416 & $17.44\pm 0.03$ & $17.35\pm 0.03$ &0.12 & 5.99 & $-0.82\pm 0.11$ & BDc & \nodata & \nodata & \nodata \\
03484769+3301185 & 57.19872 & +33.02181 & $18.65\pm 0.08$ & $17.83\pm 0.06$ &0.12 & 4.32 & $-0.83\pm 0.22$ & BDc & \nodata & \nodata & \nodata\\
\enddata
\tablecomments{The uncertainty in the spectral classification type is 0.5 subtype. BDc refers to brown dwarf candidate.}
\end{deluxetable*}
\end{rotatetable*}

\subsubsection{Physical Properties}

We infer the range of age and mass of each confirmed substellar object with the evolutionary tracks from \citet{Baraffe2015}. The bolometric magnitudes and the effective temperatures ($T_\mathrm{eff}$) for our targets were determined using the technique described in Section~6.4 of \citet{Dubber2021}. In a nutshell we used the CFHT photometry and created a grid of SpT-$A_{V}$ probability maps to calculate the bolometric magnitude and luminosity. Each point in the grid has a $\chi^{2}$ value associated, which we convert into a normalized probability. Monte Carlo analysis was performed, using 500,000 model objects and distribute proportionally across the SpT-$A_{V}$ grid using the probability for each bin. For each simulated object of each SpT-$A_{V}$ probability bin, we calculated the bolometric magnitude (using Eq~4 in \citet{Dubber2021}) by sampling Gaussian distributions of distance modulus, $d_\mathrm{mod}$ and apparent magnitude $m_J$. The Gaussian distribution of $m_J$ is created using the CFHT apparent magnitude and error. 

\begin{figure*}[htbp]
  \centering
  \includegraphics[width=0.9\textwidth]{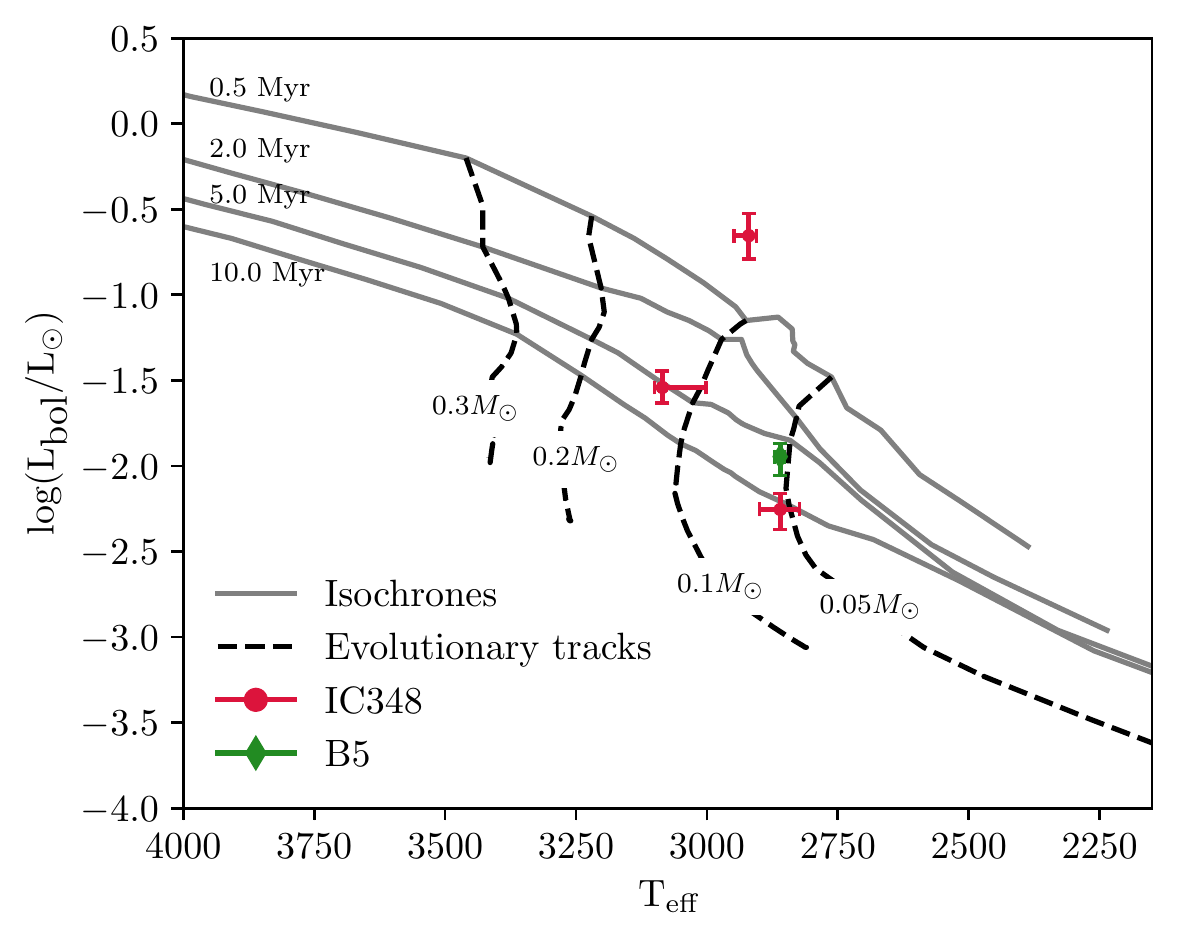}
  \caption{H-R diagram for IC\,348 (red) and B5 (green) brown dwarfs overlapped with the isochrones (solid lines; 0.5--10~Myr) and evolutionary tracks (dashed lines; 0.05~M$\sun$--0.3~M$\sun$) from \citet{Baraffe2015}, where form the mass and age are inferred.}
  \label{fig:Mass}
\end{figure*}

We adopt the bolometric correction, $BC_{J}$ for a  spectral type earlier than M7 \citep{Herczeg2015}, or for an M7 or later \citep{Filippazzo2015}. We obtain bolometric luminosity $(L_\mathrm{bol}/L_\sun)$ using Eq~5 in \citet{Dubber2021} by converting the bolometric magnitude to bolometric luminosity. We convert this grid of $(L_\mathrm{bol}/L_\sun)$ into histogram distributions, and find the peak with errors derived from 68\% Bayesian credible interval. The effective temperature ($T_\mathrm{eff}$) plotted here is estimated using the derived spectral type and relations given in Table~5 of \citet{Herczeg2014}.  Figure~\ref{fig:Mass} exhibits the results for the brown dwarfs in IC\,348 and in B5 in the H-R diagram.  

\begin{figure*}
\gridline{\fig{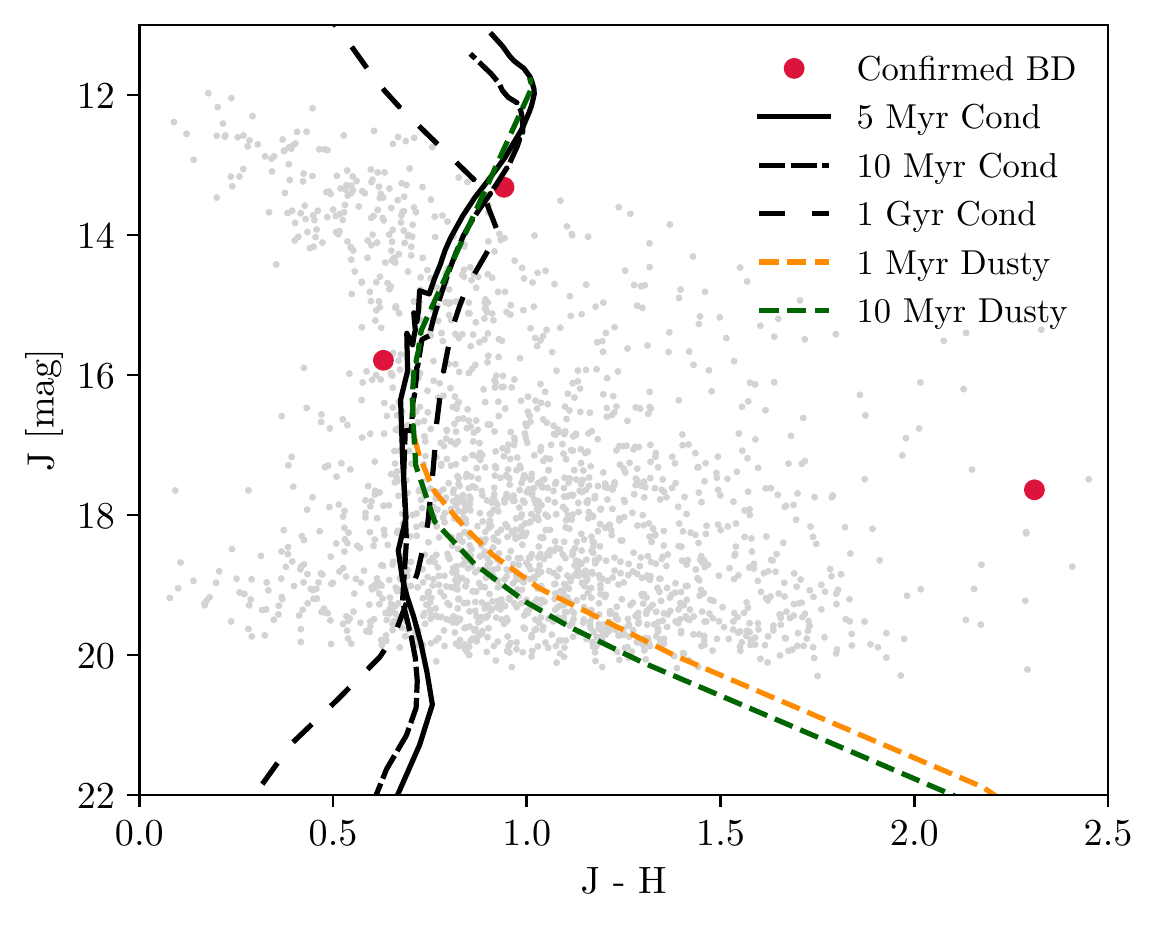}{0.70\textwidth}{(a)~IC\,348}}
\gridline{\fig{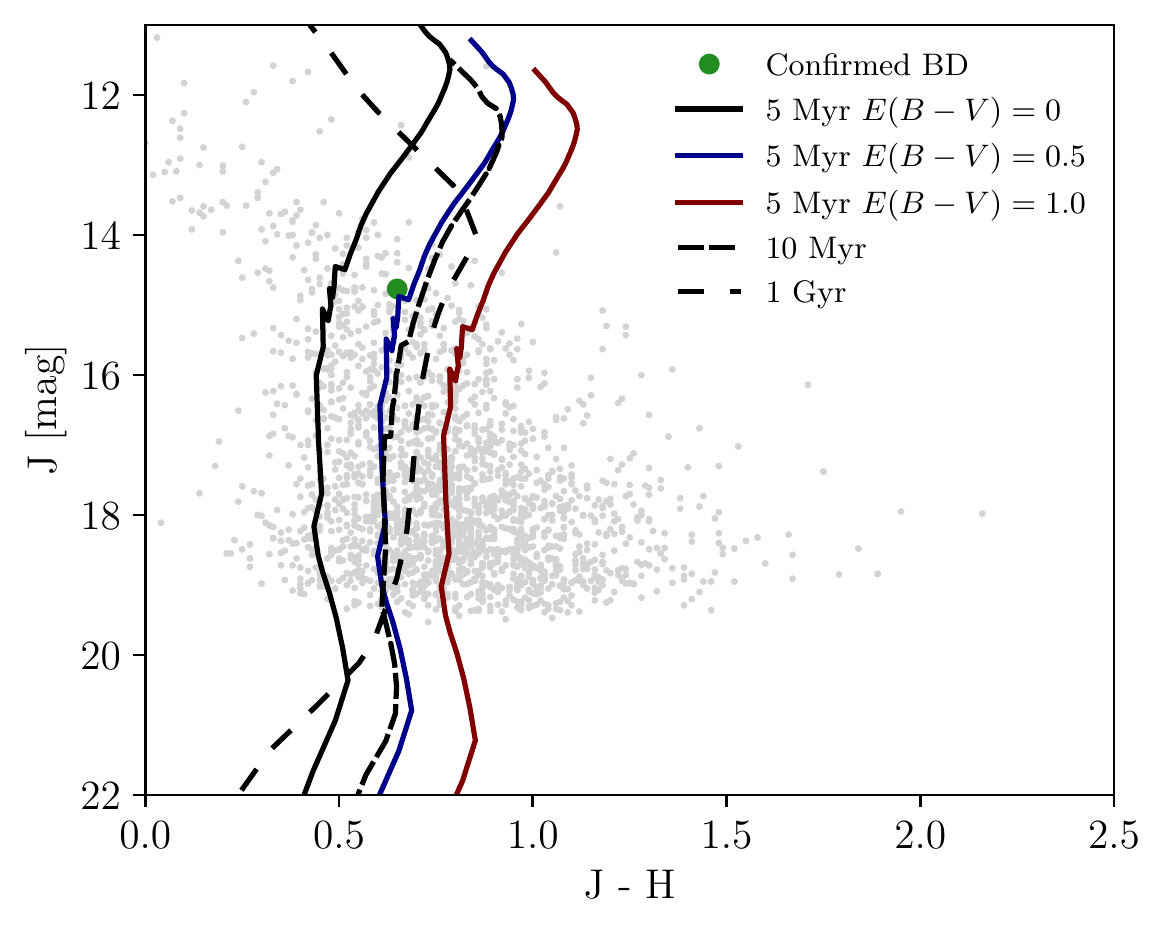}{0.70\textwidth}{(b)~B5}}
\caption{$J$ vs. $J-H$ color-magnitude diagram for (a)~IC\,348 and (b)~B5. The gray points represent all the CFHT WIRCam sources in the corresponding region whereas the confirmed brown dwarfs for IC\,348 (in red) and for B5 (in green) are marked.  Isochrones from the AMES-Cond and AMES-DUSTY model \citep{Allard2001} are overlapped, each adjusted by a distance modulus of 7.46~mag and $E(B-V)=0.71$ \citep{Trullols1997} for IC\,348, and a distance modulus of 7.72~mag and $E(B-V)=0.5$, 1.0, or 1.5 for B5, assuming $R_v=A_V/E(B-V)=3.1$, and the reddening law of $A_{J} = 0.277 A_{V}$, $A_{H} = 0.171 A_{V}$ \citep{Cardelli1989}.}
\label{fig:IC348_B5_CMD_Iso}
\end{figure*}

As a separate analysis, we constructed the color-magnitude diagram for each region, and compared the AMES-Cond and AMES-DUSTY models \citep{Allard2001} to estimate the age of the substellar objects. This is displayed in Figure~\ref{fig:IC348_B5_CMD_Iso}. In IC\,348, two out of three brown dwarfs are consistent with an age of $\sim 5$~Myr, regardless of the Cond or the Dusty model.  Here the isochrones are adjusted by a distance modulus of 7.46~mag and an extinction/reddening of $E(B-V)=0.71$~mag \citep{Trullols1997}.  In B5, the distance modulus of 7.72~mag is adopted, but the extinction to the cloud is unclear so three different values of $E(B-V)$ are exercised, at 0, 0.5, and 1.0~mag.  The single brown dwarf in B5 seems to fit well with an age of 5~Myr and a mild extinction of $E(B-V)=0.5$~mag.  

The age of IC\,348 has been estimated in the literature to be 1--6~Myr \citep{Bell2013, Cottaar2014, Luhman2003} and the three newly identified substellar objects indeed follow approximately the age range; it is assuring also that all the targets have masses close to the stellar/substellar boundary, i.e., $\lesssim 0.1~M_\sun$.

The source 03435016+3204073 in IC\,348 lies above the 0.5~Myr isochrone in Figure~\ref{fig:Mass}, therefore we cannot determine its mass or age reliably. A possible reason might be it being a reddened very low-mass star or a brown dwarf binary.  The same source appears also highly reddened in Figure~\ref{fig:IC348_B5_CMD_Iso}.  It is interesting that its $J-H$ versus $J$ location resembles that of a blackbody \citep{burrows97}.  The nature of this object deserves further delineation. 

In B5, the newly identified brown dwarf has an isochrone age of $\sim5$~Myr and a mass of $0.05~M_\sun$. This is the first discovery of a brown dwarf to date toward B5 to our knowledge. The age group assigned to this confirmed brown dwarf is INT-G ($\approx10$~Myr) which may imply that this object is likely to be a young interloper, possibly slightly closer to us than B5. Studies of such a widely distributed young low-mass objects from previous generation of newly formed stars is seen in Orion cloud complex, and is possible to see such a distribution in Perseus complex.

\subsection{Selection of Young Stellar Members In B5 Using WISE and 2MASS}

\begin{figure*}[htbp!]
\centering
\gridline{\fig{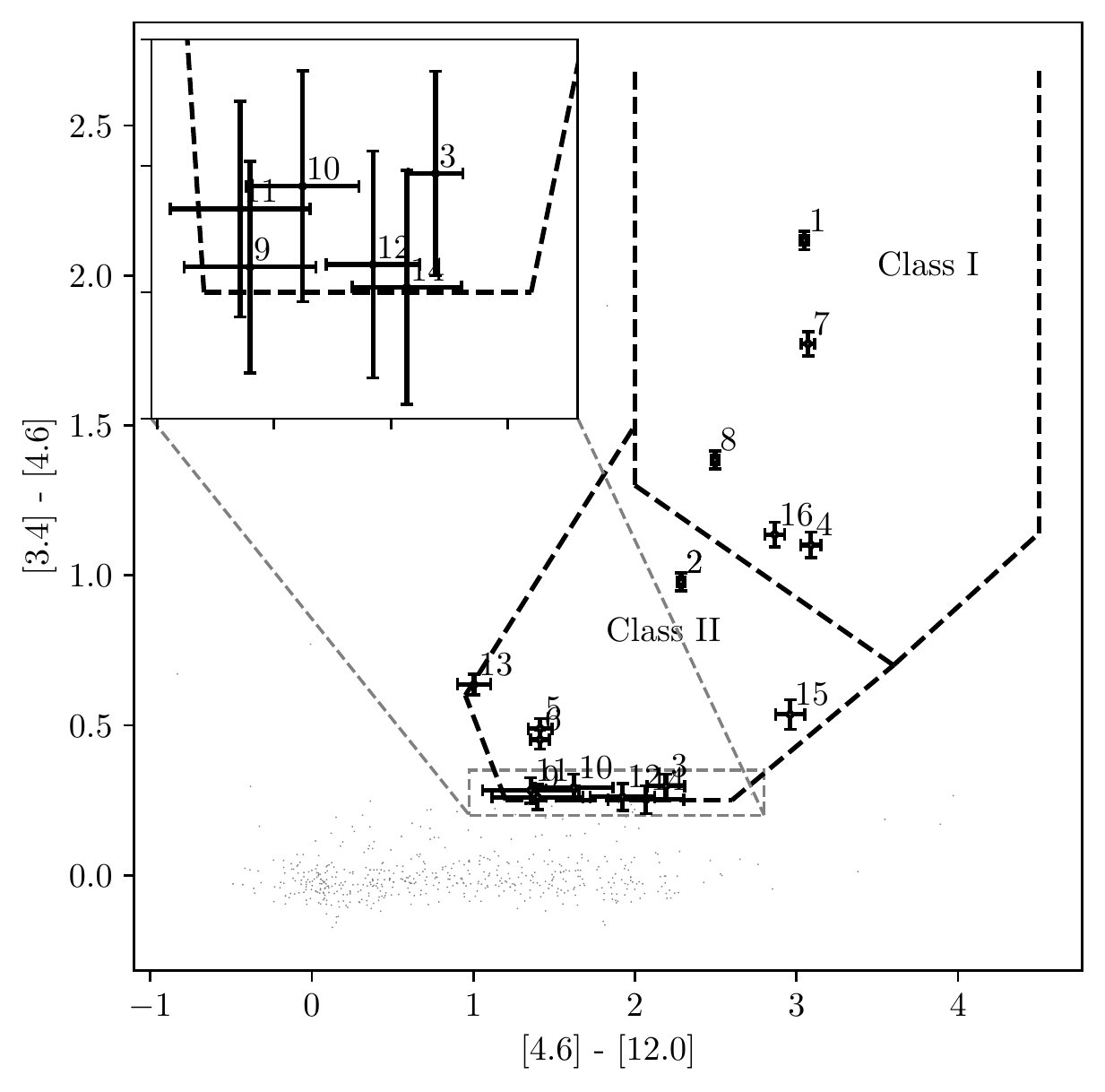}{0.6\textwidth}{(a) WISE}}
\gridline{\fig{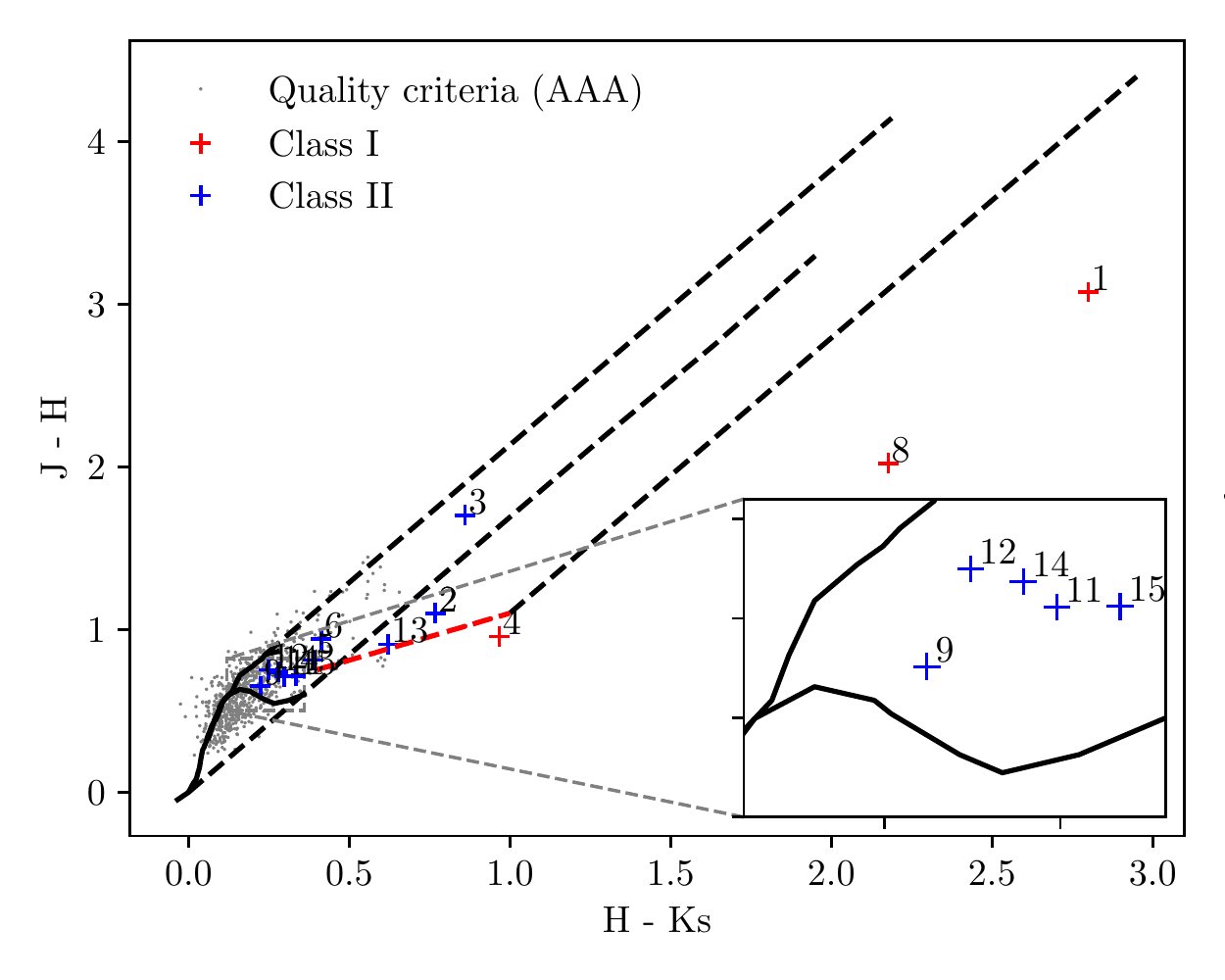}{0.65\textwidth}{(b) 2MASS}}
\caption{WISE and 2MASS color-color diagram towards B5 (a)~WISE ($W1-W2$) versus ($W2-W3$) color-color diagram to identify Class~I and Class~II objects by adopting the classification criteria given by \citet{Koenig2014}. The black dots with their respective error bars have photometric quality flag `AAA' and `AAB'. The inset shows zoom in of all the targets which lie close to the boundary of Class~II objects. (b)~2MASS ($H-Ks$) versus ($J-H$) color–color diagram. The intrinsic loci of dwarfs and giants are each marked with a black curve \citep{Bessell1988}, whereas those of classical TTauri stars are represented by the dashed red line \citep{Meyer1997}.  All the data points are transformed to the CIT system \citep{Carpenter2001}. The dashed lines represent the reddening direction from the tips of the giants, dwarfs, and TTauri loci, respectively. The Class~I and Class~II sources identified in (a) are labelled and plotted here with the same symbols. The inset shows zoom in of all the targets which lie close to the dwarf and main sequence loci.} 
\label{fig:WISE_2MASS_B5}
\end{figure*}

We adopted the scheme given by \cite{Koenig2014}, i.e., with color cuts from $W1$ to $W3$ for different YSO classes, and removing possible contaminants such as active galactic nuclei, and star-forming galaxies. Using the ($W1-W2$) versus ($W2-W3$) diagram, we identified five Class~I and 11 Class~II candidates, as shown in Fig.~\ref{fig:WISE_2MASS_B5}(a).

There are 1985 2MASS sources with quality flag `AAA'. Of the YSO candidates found by the WISE colors, Nos.~7, 16 (both Class~I) and No.~10 (Class~II) in Table~\ref{tab:B5_YSO} have no 2MASS counterparts, whereas three out of five Class~I, and 10 out of 11 Class~II sources have reliable 2MASS measurements, allowing us to conduct consistency check of their youth nature, depicted in Fig.~\ref{fig:WISE_2MASS_B5}(b). Two of the Class~II candidates, Nos.~2 and 13 in Table~\ref{tab:B5_YSO}, have consistent 2MASS colors with being T Tauri stars, whereas the rest lie along the giant loci. No.~11 as a Class~II candidate has a counterpart in our CFHT images and was selected as a $Q$ candidate for a follow-up spectroscopy. This target 03492858+3258064 turns out to be the first confirmed brown dwarf in B5.

\subsection{Kinematics of IC 348 and B5}

\begin{figure}
\centering
 \includegraphics[width=1.0\textwidth]{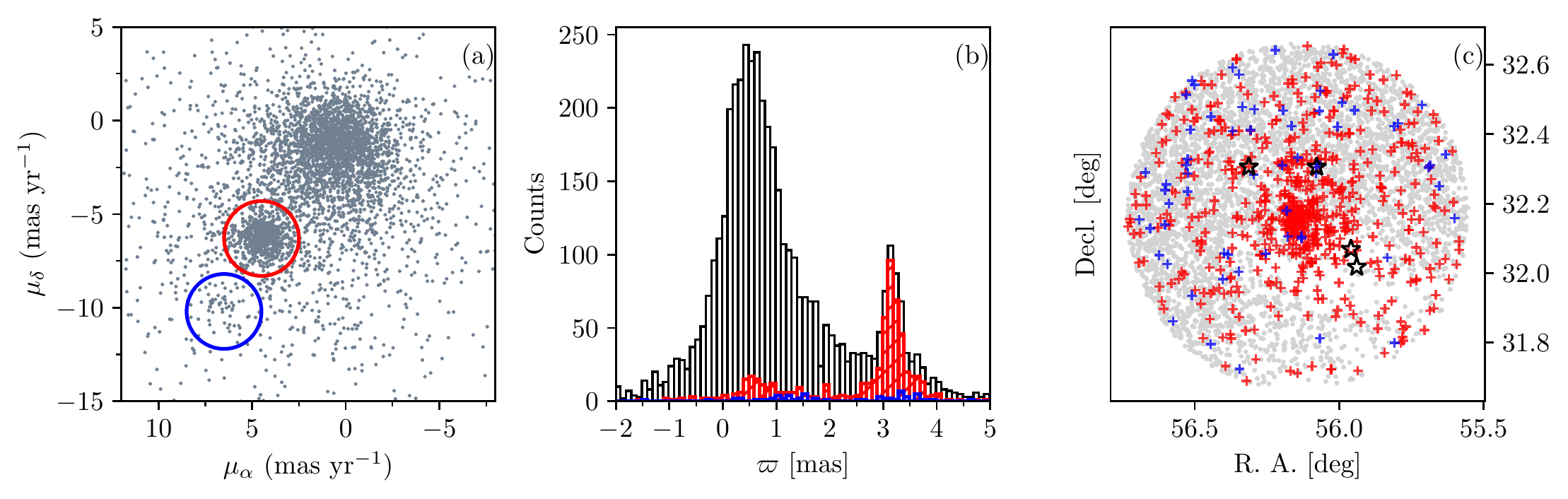}
  \caption{Gaia EDR3 $30\arcmin$ field towards IC\,348. (a) Proper-motion scatter plot (gray points) with two distinct groups marking with a red circle (Group~A) and a blue circle (Group~B), each with a radius of 2~mas~yr$^{-1}$.
  (b) Parallax distribution of the stars in (a), with Group~A shown in red and Group~B in blue. Stars belonging neither in Group~A or in Group~B, i.e., field stars, are represented in the gray histogram. (c) Sky positions of the stars in (a) using the same color scheme as in (b); that is, red: proper-motion Group~A, blue: proper-motion Group~B, whereas the gray: stars in neither group. Additionally, the black symbols represent the three confirmed substellar objects and one contaminant.
    }
  \label{fig:Gaia_IC348_subplots}
\end{figure}

The proper-motion vector plot of IC\,348 displayed in Figure~\ref{fig:Gaia_IC348_subplots}(a) shows two distinct groups, one (Group~A) centering around $\mu_{\alpha}=4.43$~mas~yr$^{-1}$, and  $\mu_{\delta}=-6.27$~mas~yr$^{-1}$ and the other (Group~B) around $\mu_{\alpha}=6.51$~mas~yr$^{-1}$, and $\mu_{\delta}=-9.99$~mas~yr$^{-1}$. The two kinematic groups are also distinguished in parallax and in sky position (Figure~\ref{fig:Gaia_IC348_subplots}(b) and \ref{fig:Gaia_IC348_subplots}(c)). Group~A is bimodal in the parallax distribution, as illustrated in Figure~\ref{fig:Gaia_IC348_subplots}(b), peaking near 0.5~mas, coincident with the field distribution, and another more populous that peaks at 3.1~mas. Group~B contains less stars, but exhibits also two maxima at about the same parallax values as above. The positional distribution clarifies the origin of each kinematic group; namely, Group~A is by and large centrally concentrated, i.e., the cluster of IC\,348 itself, whereas stars in Group~B are dispersed throughout the field. The reciprocal of the mean parallax of Group~A (3.1~mas), 323~pc is consistent with the distance to IC\,348 \citep{Ortiz-Leon2018}, confirming that Group~A is the cluster, whereas a fraction of Group~B, the ones at similar distances, is the general young stellar population permeating the Perseus complex.

\begin{figure*}[htbp!]
  \centering
  \includegraphics[width=1.0\textwidth]{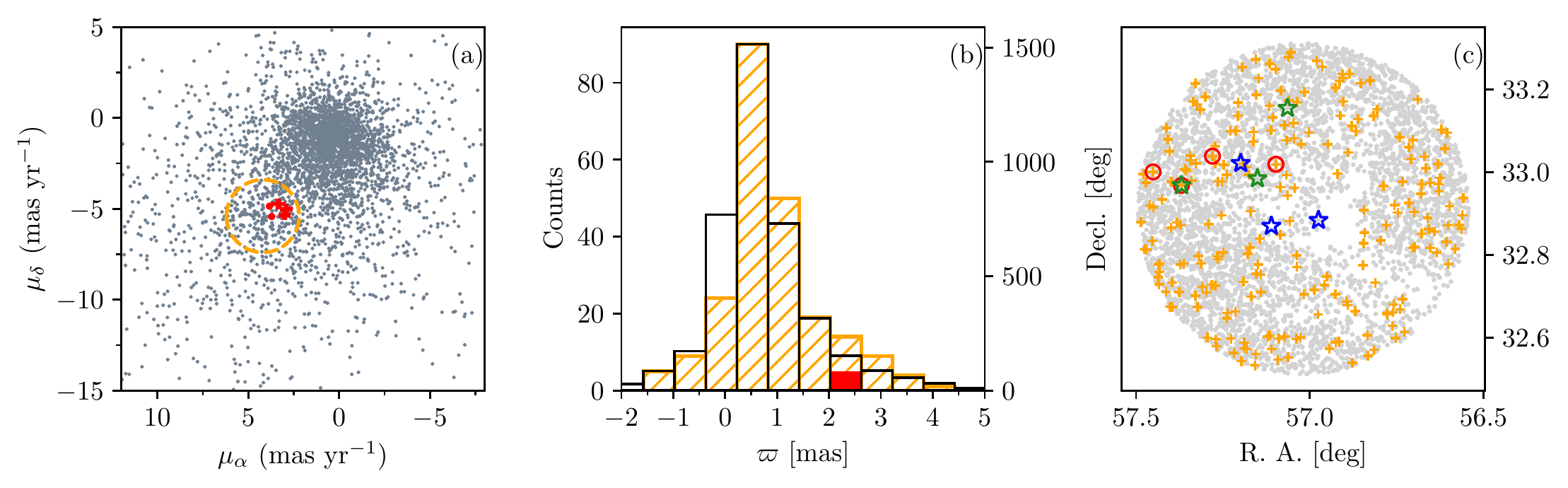}
  \caption{Same as Figure~\ref{fig:Gaia_IC348_subplots} but for Barnard~5. (a) In the proper-motion vector plot (gray points) the red symbols mark known YSOs, and the orange circle shows a radius of 2~mas~yr$^{-1}$ to encompass the possible range of proper-motion. (b) The parallax distribution of the YSOs (in red), and of the stars within the orange circle in (a) but excluding the YSOs (in orange).  Both the red and orange histograms use the ordinate scale on the left. All other stars outside the orange circle in (a) are shown as the black histogram and use the ordinate scale on the right. (c) The sky position of the three kinds of stars as in (b), with the same color scheme. Additionally, the blue symbols represent brown dwarf candidates, and the green symbols represent confirmed brown dwarf and the contaminants.}
  \label{fig:Gaia_B5_subplots}
\end{figure*}

For B5, adopting the coordinates given by \citet{Pezzuto2021}, we see no well-defined grouping in proper-motion in Figure~\ref{fig:Gaia_B5_subplots}(a), unlike that in IC\,348. We cross-matched the list of young stellar population known in the literature (i.e., Simbad) with Gaia EDR3 sources \citep{GaiaCollaboration2021}, within a coincidence radius of $5\arcsec$ resulting in 12 objects classified as ``YSO'', or a candidate ``YSO''. In the parallax distribution of the young stellar objects, there is a peak within a narrow range between 2.0~mas to 2.6~mas, (see Figure.~\ref{fig:Gaia_B5_subplots}(b)). We have listed only those YSOs that are within the given parallax range in Table.~\ref{tab:Gaia_literature}. The mean parallax, and proper-motion are, 2.35~mas, $\mu_{\alpha}=4.00$~mas~yr$^{-1}$, $\mu_{\delta}=-5.32$~mas~yr$^{-1}$ (see Table~\ref{tab:Gaia_edr3_cloud}) respectively. The reciprocal of the mean parallax corresponds to a distance of 426~pc therefore placing it farther away than what suggested by \citet{herbig1983} or \citet{Dickman1978} as 350~pc.

\section{Individual young stellar objects in B5}
\label{sec:Ind_yso_b5}

We look for additional evidence of youth by using the WISE and 2MASS colors. A total of 16 young stellar objects are found, of which one Class~I object and seven Class~II objects are newly found.  They are listed in Table~\ref{tab:B5_YSO}, in which a sequence number, the coordinates, WISE and 2MASS magnitudes, are given, followed by the name known in the literature, if any, and the classification of the object.

\movetabledown=5cm
\begin{rotatetable*}
\begin{deluxetable*}{rcc cccccc cl}
\tablecaption{\label{tab:B5_YSO} 
Young Stellar Objects Identified in Barnard~5 using WISE and 2MASS.}\tabletypesize{\scriptsize}
\tablecolumns{11}
\tablewidth{0pt}
\tablehead{\colhead{ID} &  \colhead{$\alpha_{J2000}$} & 
  \colhead{$\delta_{J2000}$} & \colhead{$W1$} & \colhead{$W2$} & \colhead{$W3$} & \colhead{$J$} & \colhead{$H$} & \colhead{$K$} & \colhead{Name\textsuperscript{a}} & \colhead{Type}\\
\colhead{} & \colhead{(deg)} & \colhead{(deg)} & \colhead{(mag)} & \colhead{(mag)} & \colhead{(mag)} & \colhead{(mag)} & \colhead{(mag)} & \colhead{(mag)} & \colhead{} & \colhead{}}
\startdata
1 & 56.92333 & +32.86216 & $8.75\pm0.02$ & $6.63\pm0.02$ & $3.58\pm0.02$ & $17.34\pm0.26$ & $14.05\pm0.04$ & $11.21\pm0.02$ & B5~IRS~1 & Class~I \\
2 & 56.94643 & +33.06756 & $8.79\pm0.02$ & $7.82\pm0.02$ & $5.53\pm0.02$ & $12.17\pm0.03$ & $11.02\pm0.03$ & $10.22\pm0.03$ & B5~IRS~4 & Class~II \\
3 & 56.90464 & +33.09548 & $13.33\pm0.03$ & $13.03\pm0.03$ & $10.84\pm0.11$ & $16.35\pm0.11$ & $14.55\pm0.05$ & $13.66\pm0.04$ & This work & Class~II \\
4 & 56.73645 & +32.88013 & $14.00\pm0.03$ & $12.89\pm0.03$ & $9.81\pm0.05$ & 17.12\nodata & $16.13\pm0.20$ & $15.12\pm0.12$ & SSTc2d~J034656.7+325248 & Class~I \\
5 & 56.73914 & +32.82148 & $11.82\pm0.02$ & $11.33\pm0.02$ & $9.92\pm0.07$ & $13.57\pm0.02$ & $12.73\pm0.02$ & $12.31\pm0.02$ & 2MASS~J03465739+3249173 & Class~II \\
6 & 56.74386 & +32.78301 & $11.81\pm0.02$ & $11.36\pm0.02$ & $9.95\pm0.06$ & $13.58\pm0.02$ & $12.61\pm0.02$ & $12.16\pm0.02$ & 2MASS~J03465851+3246588 & Class~II \\
7 & 56.74576 & +32.75514 & $13.85\pm0.03$ & $12.08\pm0.03$ & $9.01\pm0.03$ & \nodata & \nodata & \nodata & Infra-Red source & Class~I \\
8 & 56.77269 & +32.71899 & $9.60\pm0.02$ & $8.22\pm0.02$ & $5.72\pm0.02$ & $16.41\pm0.15$ & $14.26\pm0.06$ & $12.05\pm0.02$ & B5~IRS~3 & Class~I \\
9 & 57.12281 & +32.65460 & $13.40\pm0.03$ & $13.14\pm0.03$ & $11.75\pm0.28$ & $14.52\pm0.03$ & $13.86\pm0.03$ & $13.59\pm0.04$ & This work & Class~II \\
10 & 57.33036 & +32.87553 & $13.68\pm0.03$ & $13.38\pm0.04$ & $11.76\pm0.24$ & \nodata & \nodata & \nodata & This work & Class~II \\
11 & 57.36914 & +32.96846 & $13.50\pm0.03$ & $13.22\pm0.03$ & $11.86\pm0.30$ & $14.89\pm0.03$ & $14.16\pm0.04$ & $13.83\pm0.04$ & This work & Class~II \\
12 & 57.45101 & +33.00102 & $13.60\pm0.03$ & $13.34\pm0.04$ & $11.42\pm0.20$ & $14.85\pm0.03$ & $14.08\pm0.03$ & $13.80\pm0.04$ & This work & Class~II \\
13 & 56.86345 & +32.50027 & $12.25\pm0.03$ & $11.62\pm0.02$ & $10.61\pm0.10$ & $14.39\pm0.03$ & $13.45\pm0.03$ & $12.79\pm0.02$ & 2MASS~J03472722+3230010 & Class~II \\
14 & 56.87431 & +32.49139 & $13.63\pm0.03$ & $13.38\pm0.04$ & $11.31\pm0.23$ & $14.91\pm0.03$ & $14.15\pm0.04$ & $13.84\pm0.04$ & This work & Class~II \\
15 & 57.50904 & +33.08623 & $13.96\pm0.03$ & $13.43\pm0.04$ & $10.47\pm0.08$ & $15.44\pm0.05$ & $14.71\pm0.05$ & $14.34\pm0.06$ & This work & Class~II \\
16 & 57.12943 & +32.43432 & $13.66\pm0.03$ & $12.53\pm0.03$ & $9.66\pm0.06$ & \nodata & \nodata & \nodata & This work & Class~I
\enddata
\tablecomments{a. Please see Sec.~\ref{sec:Ind_yso_b5} for further details about the individual targets.}
\end{deluxetable*}
\end{rotatetable*}

\movetabledown=5cm
\begin{rotatetable*}
\begin{deluxetable*}{lcccccccl}
\tablecaption{\label{tab:Gaia_literature} Gaia EDR3 Data of Young Stellar Objects.}
\tablecolumns{9}
\tablehead{\colhead{$\alpha_{J2000}$} & \colhead{$\delta_{J2000}$} & \colhead{$\varpi$} & \colhead{$\mu_{\alpha}$} & \colhead{$\mu_{\delta}$} & \colhead{$G$} & \colhead{$G_{\rm BP}$} &\colhead{$G_{\rm RP}$} & \colhead{Name}\\
\colhead{(deg)} & \colhead{(deg)} & \colhead{(mas)} & \colhead{(mas~yr$^{-1}$)} & \colhead{(mas~yr$^{-1}$)} & \colhead{(mag)} & \colhead{(mag)} & \colhead{(mag)} & \colhead{}}
\startdata
57.0973 & +33.0194 & $2.41\pm0.06$ & $3.71\pm0.09$ & $-5.42\pm0.05$ & $15.73\pm0.004$ & $17.44\pm0.02$ & $14.33\pm0.01$ & Candidate YSO \citep{Zari2018}\\
57.2800 & +33.0390 & $2.30\pm0.08$ & $3.84\pm0.10$ & $-4.86\pm0.07$ & $16.90\pm0.003$ & $19.29\pm0.03$ & $15.50\pm0.01$ & Candidate YSO \citep{Zari2018} \\
57.3692 & +32.9684 & $2.47\pm0.26$ & $3.35\pm0.33$ & $-4.73\pm0.21$ & $18.99\pm0.003$ & $21.33\pm0.11$ & $17.52\pm0.01$ & Source No.~11\\
57.0977 & +33.0193 & $2.61\pm0.14$ & $2.91\pm0.28$ & $-5.03\pm0.25$ & \nodata & \nodata & \nodata & Candidate YSO \citep{Zari2018} \\
57.4510 & +33.000 & $2.62\pm0.21$ & $3.00\pm0.27$ & $-5.36\pm0.17$ & $18.61\pm0.003$ & $20.86\pm0.13$ & $17.17\pm0.01$ & Source No.~12\\ \hline
\multicolumn{9}{c}{\textbf{WISE and 2MASS Targets Gaia Counterparts}} \\
56.9464 & +33.0675 & $4.77\pm0.29$ & $1.21\pm0.40$ & $-4.17\pm0.22$ & $15.77\pm0.01$ & $16.93\pm0.02$ & $14.32\pm0.02$ & B5~IRS~4 \\
56.7391 & +32.8215 & \nodata & \nodata & \nodata & $18.34\pm0.01$ & $19.67\pm0.04$ & $16.08\pm0.01$ & Source No.~5\textsuperscript{*}  \\
56.7439 & +32.7830 & \nodata & \nodata & \nodata & $17.88\pm0.01$ & $20.39\pm0.07$ & $16.27\pm0.01$ & Source No.~6\textsuperscript{*} \\
57.1229 & +32.6545 & $7.97\pm0.14$ & $73.70\pm0.16$ & $-5.73\pm0.10$ & $17.77\pm0.003$ & $19.65\pm0.04$ & $16.46\pm0.01$ & Source No.~9\textsuperscript{*} \\
57.3304 & +32.8755 & $15.28\pm0.25$ & $-64.27\pm0.32$ & $-337.94\pm0.19$ & $18.62\pm0.003$ & $21.44\pm0.12$ & $17.12\pm0.01$ & Source No.~10\textsuperscript{*} \\
56.8635 & +32.5002 & $2.24\pm0.34$ & $3.52\pm0.40$ & $-5.76\pm0.28$ & $19.40\pm0.02$ & 	$21.03\pm0.11$ & $17.85\pm0.06$	& Source No.~13\textsuperscript{*} \\
56.8743 & +32.4914 & $3.56\pm0.17$ & $23.87\pm0.19$ & $-8.07\pm0.14$ & $18.45\pm0.003$ & $20.43\pm0.08$ & $17.07\pm0.01$ & Source No.~14$^{\dagger}$ \\
57.5090 & +33.0862 & $2.57\pm0.32$ & $3.05\pm0.37$ & $-4.56\pm0.28$ & $19.42\pm0.004$ & $21.43\pm0.14$ & $17.96\pm0.01$ & Source No.~15$^{\dagger}$ \\
57.1294 & +32.4343 & $0.031\pm0.28$ & $-0.36\pm0.31$ & $-0.32\pm0.24$ & $19.10\pm0.01$ & $19.46\pm0.03$ & $18.32\pm0.02$ & Source No.~16\textsuperscript{*}
\enddata
\tablecomments{\begin{itemize}
    \item[*] These sources have Gaia counterparts either with no proper-motion data or with a large error.
    \item[$\dagger$] Source No.~14 and Source No.~15 are slightly outside our Gaia field, but both are within our proper-motion criterion.
\end{itemize}
}
\end{deluxetable*}
\end{rotatetable*}

\citet{beichman1984} reported four IRAS sources in the vicinity of B5, among which IRS~1 (IRAS~$03445+3242$) is the youngest and the first embedded YSO. Located near the center of the cloud, the source (No.~1 in Table~\ref{tab:B5_YSO}) is very cold, temperature and luminosity being 30--100~K, 10~L$_\sun$ respectively, \citet{beichman1984} and are associated with CO outflows \citep{Zapata2014}, bow-shocks, Herbig-Haro objects \citep[HH~366E and HH~366W,][]{Yu1999, bally1996}, and shock-excited H$_{2}$ emissions. IRS~1 has a circumstellar disk whose orientation is perpendicular to the CO outflow \citep{ Langer1996, fuller1991}, and also exhibits a highly collimated optical jet \citep{bally1996}. NH$_{3}$(1,1) and (2,2) observations with the JVLA \citep{Pineda2015} additionally revealed the presence of filaments within the quiescent dense core. These filaments enclose three high-density bound condensations which, combined with IRS~1, might form a bound quadruple multiple star system \citep{Pineda2015}. The WISE and 2MASS colors of IRS~1 suggests all consistence a Class~I object.

Source No.~2 in (Table~\ref{tab:B5_YSO}) is IRS~4 (IRAS~$03446+3254$), to the northeast of B5 and is the most evolved known YSO in B5 \citep{goldsmith1986,langer1989,Yu1999}. Molecular line maps show a large cavity having a diameter of $6\arcmin$ \citep{goldsmith1986,langer1989,Arce2011}, likely tracing a much older outflow. This target has a WISE Class~II colors and a 2MASS colors consistent with being a mildly reddened classical TTauri star.

Source No.~3 previously not identified in the literature. We have classified it as a possible Class~II, or a reddened dwarf.

Source No.~4, SSTc2d~J034656.7+325248 was identified by \citet{Dunham2008} as a low-luminosity Class~I protostar with 0.015~L$_\sun$ based on 3.6--70~\micron\ Spitzer data. This target is a WISE Class\,I, and is very close to the 2MASS classical TTauri loci.

Both Source No.~5 and 6 were identified as Class~II YSOs by \citet{Dunham2015, Young2015, Evans2009} based on 3.6--24~\micron\ Spitzer data, with the extinction corrected luminosity of 0.039~L$_\sun$ (source no.~5) and 0.047~L$_\sun$ (source no.~6). In Figure~\ref{fig:WISE_2MASS_B5}(a) both sources appear close to each other, and in 2MASS both the target appear together at the edge of the classical TTauri loci.

Source No.~7 is a known infrared source. The source does not have a 2MASS counterpart with the WISE colors suggesting a reddened Class~I YSO, though \citet{Hsieh2013} based on SPITZER c2d data have classified it as a probable galaxy contaminant. 
Source No.~8 is IRS~3 (IRAS~$03439+3233$) located towards the southwest of IRS~1, and drives small-scale Herbig-Haro objects, HH~367A and HH~367B, with HH~844 extending to the south \citep{Walawender2005a}. This target is identified as a Class~I object judging by both WISE and 2MASS colors.

Source No.~9 is classified marginally as a WISE Class~II source, but in 2MASS the target is reddened and lies in the dwarf loci. Source No.~10 is also a marginal WISE Class~II source. It does not have a 2MASS counterpart, and if reddening is considered, it could still be a Class~II.

Source No.~11 is identified as Class~II object and has a counterpart in our CFHT images. This target was followed up spectroscopically as a $Q$ candidate brown dwarf in B5. In the 2MASS color-color plot this target lies near the dwarf loci. Source No.~12 is classified as Class~II object based on WISE colors, but in 2MASS the target lies in the main sequence loci.

Source No.~13 is known as 2MASS~J03472722+3230010. It is identified by \citet{Yao2018} as a disk-bearing source in IC\,348. Considering its sky position, we associate it with B5 instead, and classify it as a Class~II object.

Source No.~14 is classified as a WISE Class~II object. Source No.~15 appears to be more reddened than source No.~14. In 2MASS both the targets appear in the reddened dwarf loci.  Source No.~16 is a WISE Class~I object, but does not have a 2MASS counterpart.

Note that of the four IRAS sources originally reported by \citet{beichman1984}, IRS~2 (IRAS~$03449+3240$) is colder (color temperature $\sim25$~K) but much less luminous (1.3~L$_\sun$) than IRS~1. It was supposed to be either an internally powered prestellar core or an externally heated clump \citep{beichman1984}. This source is not included in our YSO list because it is highly reddened ($W2-W3$)=4.8, and therefore not categorized as any YSO class.

\section{Bulk motion of Perseus subclusters}
\begin{figure}[h!]
\gridline{\fig{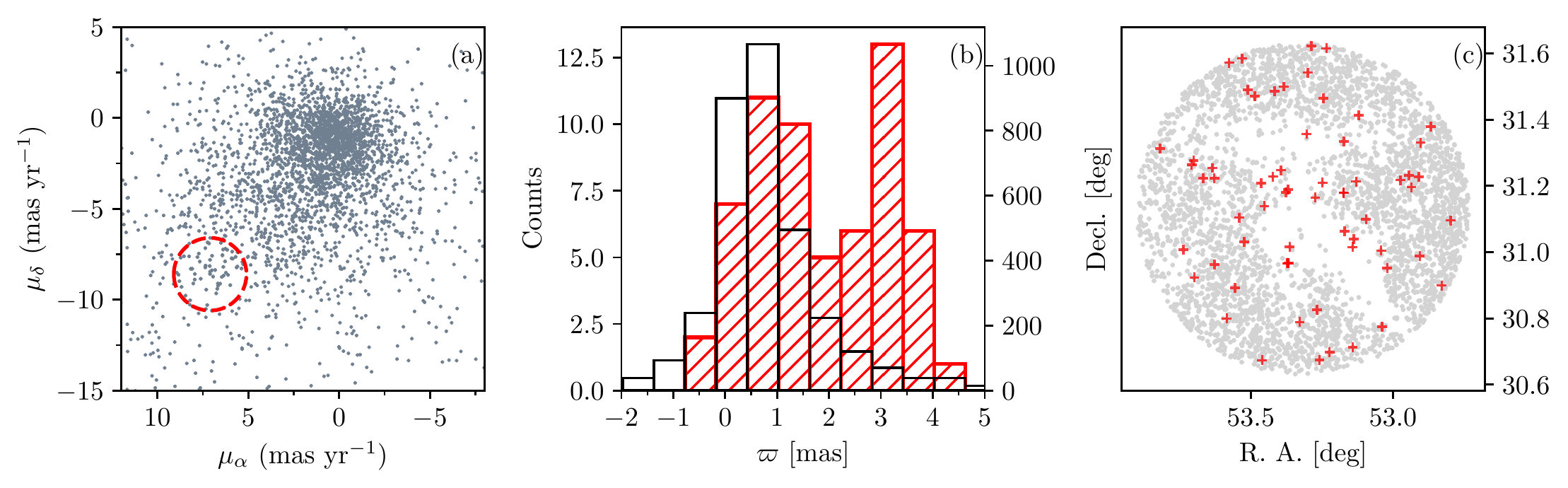}{1.0\columnwidth}{(a) Barnard~1}}
\gridline{\fig{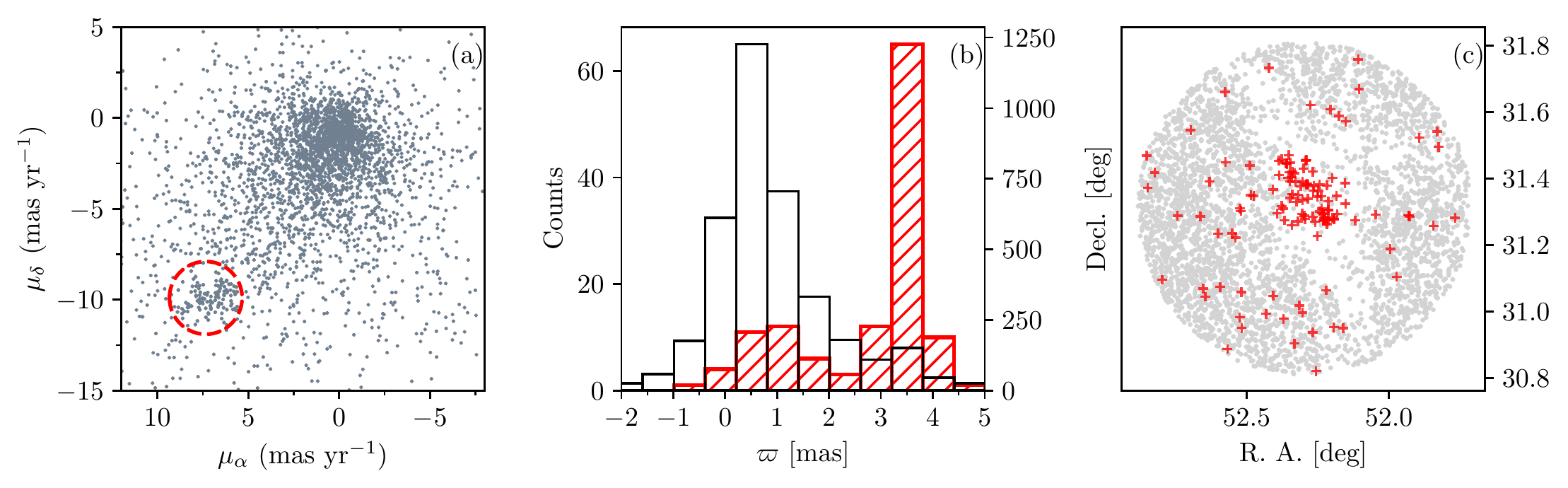}{1.0\columnwidth}{(b) NGC\,1333}}
\gridline{\fig{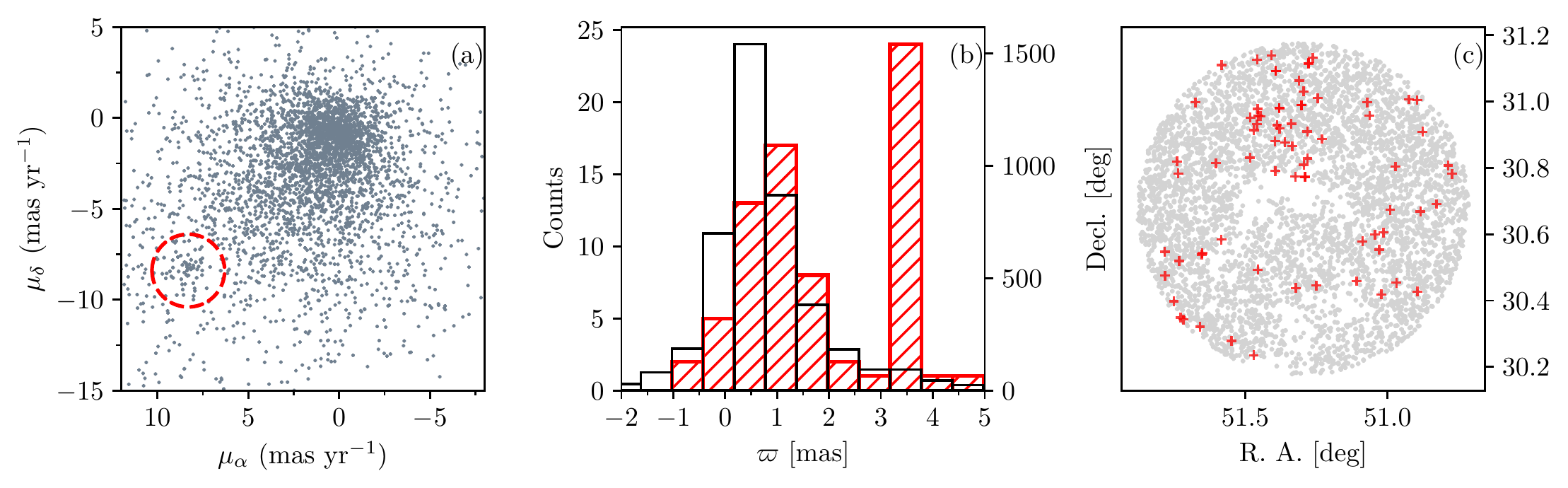}{1.0\columnwidth}{(c) L~1448}}
\caption{Same as Figure~\ref{fig:Gaia_IC348_subplots} but for the three star-forming regions of Perseus molecular cloud.}
\label{fig:Gaia_3regions}
\end{figure}

In addition to the study of B5 and IC\,348 detailed above, we consider with Gaia EDR3 data of three other star-forming regions, namely, continuing westwards, Barnard~1, NGC~1333, and L~1448 (Barnard~203), together spanning across the Perseus molecular cloud.  As in the analysis of IC\,348, the kinematic grouping of each region is utilized to find the average proper-motion and parallax. For each of the regions we analyzed the Gaia EDR3 data in a $\sim30\arcmin$ Gaia EDR3 field (see Figure~\ref{fig:B5_IC348_CO}). This angular size of the field was not meant to be optimal, but considered sufficiently large so as to cover the entire region to recognize a reliable group of stars in the proper-motion distribution. In B5, no such distinct kinematic grouping is obvious, so a sample of young stars were utilized, the membership sample is small and within a narrow range of parallax (cf.~Figure~\ref{fig:Gaia_B5_subplots}). Each of the remaining four regions exhibits a separate proper-motion grouping for which the average proper-motion and parallax are computed. In every case a secondary peak exists that mimics the shape of the field distribution. The results of the Gaia EDR3 data for Barnard\,1, NGC\,1333, and L\,1448 are presented in Figure~\ref{fig:Gaia_3regions}.

The IC\,348 cluster experiences relatively low extinction, but is surrounded by heavy filaments, delineated by the paucity of Gaia EDR3 stars, in the south where ongoing star formation is taking place (cf., Figure~\ref{fig:Gaia_IC348_subplots}(c)). The other rich cluster NGC\,1333 in contrast coincides with the dense parts of the cloud, and has a shape elongated in NE-SW.  L\,1448 on the other hand seems to be offset to the north from the obscured patches.  Barnard\,1 appears marginally within our default 30\arcmin\ field, with the proper-motion and parallax ``candidate members'' scattering throughout the field, with no spatial concentration. Gaia star counts reveal the same NW-SW dark filament direction as detected by \citet{Walawender2005a}.

\citet{Ortiz-Leon2018} reported detailed VLBA and Gaia DR2 astrometry for IC\,348 and NGC\,1333, and with mean radial velocities, derived the 3D motion of the two clusters. For IC\,348, our estimates of the average parallax of $\varpi=3.14 \pm 0.1$~mas, and
of the proper-motion of $4.4 \pm 0.8, -6.3 \pm 0.7$~mas~yr$^{-1}$ agree within uncertainties with their Gaia parallax of $\varpi_{\rm Gaia}=3.09$~mas, VLBA parallax of $\varpi_{\rm VLBA}=3.12$~mas, and proper-motion of $4.35, -6.76$~mas~yr$^{-1}$. For NGC\,1333, our values are also consistent with those reported by \citet{Ortiz-Leon2018} based on Gaia DR2 measurements.

\begin{figure}[h!]
  \centering
   \includegraphics[width=\columnwidth]{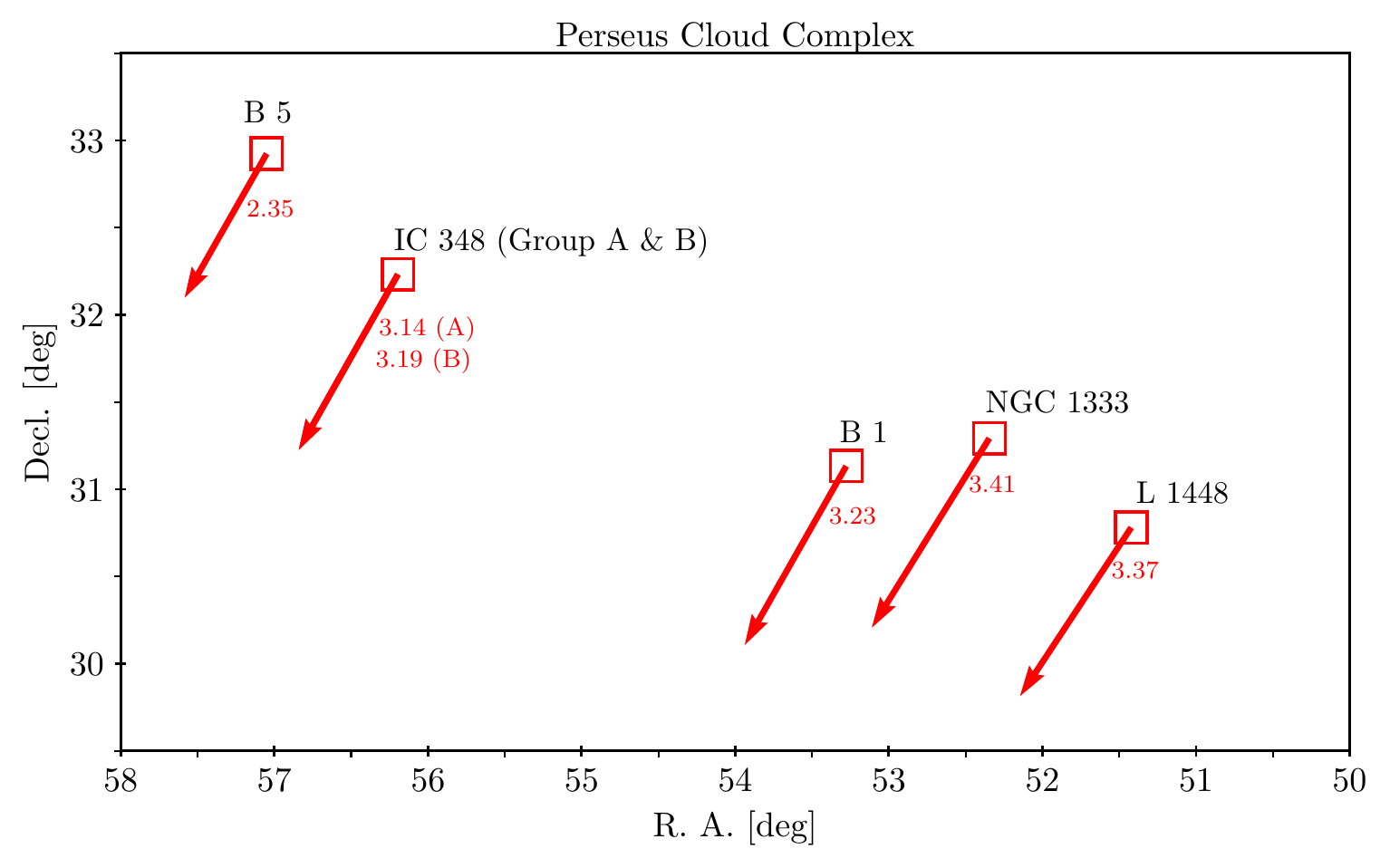}
  \caption{Proper-motion and parallax distribution of the selected regions in the Perseus complex. Each red square marks the sky position of a region with an arrow illustrating the relative proper-motion (not to scale with the coordinate axes). The number next to the square denotes the average of the parallax values (given in mas).}
 \label{fig:perseus}
\end{figure}

The overall motion and orientation of the Perseus cloud complex probed by the five selected regions are summarized in Table~\ref{tab:Gaia_edr3_cloud} and depicted in Figure~\ref{fig:perseus}.  In our analysis we compared the parallax measurement of each cloud, rather than the distance itself even though at this close distance the bias introduced in the reciprocal function in estimate of the distance \citep{bai21} is expected to be minimal. Our results support the earlier notion that the western complex is closer than the eastern part \citep{Schlafly2014, Ortiz-Leon2018}, and with a systematic trend in both proper-motion and parallax extending further to include B5, the easternmost edge of the complex.  While the near side is at some 300--320~pc \citep{Pezzuto2021}, B5 stands alone to be $\sim100$~pc farther away.  

\begin{deluxetable*}{lCCCCCC}
\tablecaption{\label{tab:Gaia_edr3_cloud} Proper-motion and Parallax of Regions of Perseus molecular Cloud.}
\tablecolumns{7}
\tablehead{\colhead{Region} & \colhead{$\alpha_{J2000}$} & \colhead{$\delta_{J2000}$} & \colhead{$\mu_{\alpha}$} & \colhead{$\mu_{\delta}$} & \colhead{$\varpi$} & \colhead{$1/\varpi$}\\
\colhead{} & \colhead{(deg)} & \colhead{(deg)} & \colhead{(mas~yr$^{-1}$)} & \colhead{(mas~yr$^{-1}$)} & \colhead{(mas)} & \colhead{pc}}
\startdata
Barnard\,5         & 56.90 & 32.90 & 4.00\pm0.94  & -5.32\pm0.67  & 2.35\pm0.76 & $426\pm32$ \\
IC\,348 (Group~A) & 56.13 & 32.16 & 4.43\pm0.20   & -6.27\pm0.14  & 3.14\pm0.18 & $318\pm37$ \\
IC\,348 (Group~B) & 56.13 & 32.16 & 6.51\pm0.28   & -9.99\pm0.21  & 3.19\pm0.23 & $313\pm47$ \\
Barnard\,1         & 53.32 & 31.13 & 7.03\pm0.16  & -8.31\pm0.12  & 3.23\pm0.14 & $311\pm40$ \\
NGC\,1333         & 52.30 & 31.31 & 7.10\pm0.17   & -9.92\pm0.12  & 3.41\pm0.14 & $292\pm47$ \\
L\,1448            & 51.29 & 30.68 & 8.39\pm0.10  & -8.22\pm0.07  & 3.37\pm0.08 & $297\pm62$ \\
\enddata
\end{deluxetable*}

For B5, \citet{langer1989} concluded with the optically thin $^{13}$CO tracer to derive some 1000~M$_\sun$ of gas and dust, comparable to the virial mass, with a  relative small fraction ($\sim200~M_\sun$) in the core and fragments. Our YSO census within the $30\arcmin$ field using 2MASS and WISE should be reasonably complete down to the magnitude limits. Our $W$ band image though limited in spatial coverage led to the encouraging discovery of one brown dwarf. A survey more comprehensive than we reported here for the entire B5 cloud is desirable to further add to the substellar sample. In any case, given the few known YSOs, of which many are Class~I objects, B5 appears to be in infancy in star formation and low in formation efficiency. The cloud subtends $\sim1\arcdeg$ in the sky \citep{Bensch2006} which, adopting a distance of 400~pc, corresponds to a 7~pc linear dimension. B5 is certainly connected with the complex, witnessed in the continuation of the distribution of CO (see Figure~\ref{fig:B5_IC348_CO}) or the Herschel far-infrared data \citep[their Fig.~4][]{Ortiz-Leon2018}. If so, the fact that B5 stretches a depth (100~pc) more than the length scale projected on the sky plane suggests a bending morphology, alongside or on the backside of Per~OB2 bubble, of this part of the complex.  

Massive stars upon birth interact and impact the neighboring environments. They may disrupt a nearby cloud to hamper any star formation activity. At a distance, however, these luminous stars via their winds and radiation could instead play a constructive role in compressing a cloud, thereby prompting the formation of the next generation of stars \citep{lee2005,chen2007,lee2007}. A supernova explosion has an even longer-range effect and produces a cavity that plows into an ever denser expanding shell of gas and dust, within which gravitational instability then leads to formation of giant molecular clouds that fragment hierarchically to form stellar groups \citep{McCray1987}. Such a ``collect-and-collapse' process can propagate on large ($> 100$~pc) scales, and indeed embedded star clusters are detected at the peripheral shells/rings of some \ion{H}{2} regions \citep{zavagno2006}.  Differing timing of the explosive or photoionization shock fronts encountering with the surrounding clouds, plus possibly multiple supernova events in an OB association, results in an age spread in the YSOs thus induced to form.  It has been suggested that Perseus OB2 (Per~OB2) has such an influence in triggering the star formation in the Perseus cloud complex.  

Per~OB2, located at a distance of $\sim 300$~pc, is one of the major OB associations in the solar vicinity \citep{deZeeuw1999}.  Supernova explosion in Per OB2 drives an expanding \ion{H}{1} supershell with a diameter of $\sim 20\arcdeg$ into the surrounding interstellar medium \citep{Sancisi1974, Heiles1984, Hartmann1997}. The eastern end of the Perseus cloud is close to the centroid of the supershell and to the OB association, exposing the part of the cloud now the cluster IC\,348 resides as the first contact some 6~Myr ago in the sequence of triggered star formation \citep{deZeeuw1999, Ridge2006a, Bally2008}.  

Comparatively, the cluster NGC\,1333 together with the associated reflection nebula and dark cloud L\,1450 (Barnard\,205) is the most rigorous star-forming region in the Perseus complex, manifest amply with embedded smm cores, protostellar jets/outflows, and  shock-excited HH objects.  With an estimated age of $\sim1$~Myr, the total stellar mass of $\sim100$~M$_\sun$ signifies a star formation rate of $10^{-4}~M_\sun$~yr$^{-1}$ \citep{Walawender2008}. Except for the cluster IC\,348 representing the earlier epoch of starbirth, the current episode of star formation appears to pervade throughout the complex including the regions we present in this work.


\section{Conclusion}

We present results from the $W$-band survey to search for brown dwarfs in IC\,348 and in B5, and broad-band infrared colors to find young stars in B5.  Our main results are:

\begin{enumerate}
  \item $W$-band imaging, devised to detect the water absorption feature commonly seen in substellar objects, continues to demonstrate its effectiveness, providing a high positive rate of 90\% for follow-up confirmation spectroscopy, to filter out reddened background sources. This work applied this technique to two clouds at the eastern edge of the Perseus star-forming complex.  
  
  \item We found three previously unknown substellar objects in IC\,348, enhancing the substellar sample in addition to the relatively well known stellar membership in the cluster.

  \item Barnard\,5 appears to be isolated with low level of star-forming activity. Our $W$-band images covering part of the cloud led to the discovery of the first brown dwarf in this region. Using WISE and 2MASS data we identified 16 young stellar objects, of which one new Class~I and seven Class~II objects were previously unknown. Barnard\,5 should be at the initial stage of star birth, not necessarily as influenced by the neighboring Per~OB2 as is the case for IC\,348.
     
  \item Using the Gaia EDR3 data of five selected star-forming regions in the complex, we derived the distances of L\,1448, NGC\,1333, Barnard\,1, and IC\,348, progressively more distant from southwest to northeast, ranging 300--320~pc. The Barnard\,5 cloud, however, is found to be considerably farther at $\sim426$~pc.
 
\end{enumerate}

\vspace{5mm}

\software{\texttt{astropy} \citep{2013A&A...558A..33A,2018AJ....156..123A}}
          
\section{ACKNOWLEDGEMENTS}

The authors B.L. and W.-P.C. acknowledge financial support from the MOST grant 109-2112-M-008-015-MY3 to carry out this study. Based on observations obtained with WIRCam, a joint project of CFHT, Taiwan, Korea, Canada, and France, at the Canada–France–Hawaii Telescope (CFHT), which is operated by the National Research Council (NRC) of Canada, the Institut National des Sciences de l’Univers of the Centre National de la Recherche Scientifique of France, and the University of Hawaii. Micka\"{e}l Bonnefoy and Philippe Delorme acknowledge support in France from the French National Research Agency (ANR) through project grant ANR-20-CE31-0012 and the Programmes Nationaux de Planetologie et de Physique Stellaire (PNP \& PNPS). B.L. acknowledges the support of Bobby Bus for the help and guidance on observing with IRTF SpeX.
This research has benefited from the SpeX Prism Spectral Libraries, maintained by Adam Burgasser at \url{http://pono.ucsd.edu/~adam/browndwarfs/spexprism/}. This publication makes use of data products from the Wide-field Infrared Survey Explorer, which is a joint project of the University of California, Los Angeles, and the Jet Propulsion Laboratory/California Institute of Technology, funded by the National Aeronautics and Space Administration.
This publication makes use of data products from the Two Micron All Sky Survey, which is a joint project of the University of Massachusetts and the Infrared Processing and Analysis Center/California Institute of Technology, funded by the National Aeronautics and Space Administration and the National Science Foundation. This publication makes use of data products from the Wide-field Infrared Survey Explorer, which is a joint project of the University of California, Los Angeles, and the Jet Propulsion Laboratory/California Institute of Technology, funded by the National Aeronautics and Space Administration. 
This work has made use of data from the European Space Agency (ESA) mission Gaia (\url{https://www.cosmos.esa.int/gaia}), processed by the Gaia Data Processing and Analysis Consortium (DPAC, \url{https://www.cosmos.esa.int/web/gaia/dpac/consortium}). This research has made use of the services of the ESO Science Archive Facility. This research has made use of the SIMBAD database, operated at CDS, Strasbourg, France, and NASA’s Astrophysics Data System.

\bibliography{wbis_ic348}{}
\bibliographystyle{aasjournal}

\end{document}